\documentclass[journal=jctcce,manuscript=article]{achemso}

\usepackage{chemformula} 
\usepackage[T1]{fontenc} 
\usepackage{graphicx}
\usepackage{float}
\usepackage{amsmath}
\usepackage{amssymb}
\usepackage{subfigure}
\usepackage{braket}
\usepackage{afterpage}
\usepackage{diagbox}
\usepackage{multirow}
\usepackage{xr}
\definecolor{ao(english)}{rgb}{0.0, 0.42, 0.24}

\author{Xinyuan Liang}
\affiliation{Academy for Advanced Interdisciplinary Studies, Peking University, Beijing, 90871, P. R. China}
\alsoaffiliation{HEDPS, CAPT, College of Engineering, Peking University, Beijing, 100871, P. R. China}
\author{Renxi Liu}
\affiliation{Academy for Advanced Interdisciplinary Studies, Peking University, Beijing, 90871, P. R. China}
\alsoaffiliation{HEDPS, CAPT, College of Engineering, Peking University, Beijing, 100871, P. R. China}
\author{Mohan Chen}
\email{mohanchen@pku.edu.cn}
\affiliation{AI for Science Institute, Beijing 100080, P. R. China}
\alsoaffiliation{Academy for Advanced Interdisciplinary Studies, Peking University, Beijing, 90871, P. R. China}
\alsoaffiliation{HEDPS, CAPT, College of Engineering, Peking University, Beijing, 100871, P. R. China}

\title{A Deep Learning Framework for the Electronic Structure of Water: Towards a Universal Model}

\begin{document}




\begin{abstract}
Accurately modeling the electronic structure of water across scales, from individual molecules to bulk liquid, remains a grand challenge. 
Traditional computational methods face a critical trade-off between computational cost and efficiency.
We present an enhanced machine-learning Deep Kohn-Sham (DeePKS) method for improved electronic structure, DeePKS-ES, that overcomes this dilemma. 
By incorporating the Hamiltonian matrix and their eigenvalues and eigenvectors into the loss function, we establish a universal model for water systems, which can reproduce high-level hybrid functional (HSE06) electronic properties from inexpensive generalized gradient approximation (PBE) calculations.
Validated across molecular clusters and liquid-phase simulations, our approach reliably predicts key electronic structure properties such as band gaps and density of states, as well as total energy and atomic forces.
This work bridges quantum-mechanical precision with scalable computation, offering transformative opportunities for modeling aqueous systems in catalysis, climate science, and energy storage.
\end{abstract}


\section{Introduction}

Water, the ubiquitous solvent of life and a cornerstone of countless chemical, biological, and environmental processes, drives persistent efforts to resolve its microscopic properties.
First-principles simulations based on density functional theory (DFT)~\cite{64PR-Hohenberg,65PR-Kohn} have been widely used to probe the multiscale behavior of water, including microscopic structures of molecular clusters to liquid water,~\cite{14JCP-DiStasio,17PNAS-Chen} hydrated ions~\cite{02Nature-Tuckerman} and ice.~\cite{11PRL-Santra}
While generalized gradient approximations (GGAs)~\cite{77PRL-Perdew} enable practical simulations of aqueous systems, higher-tier approaches, particularly hybrid functionals~\cite{96JCP-Perdew} and many-body methods like the random-phase approximation (RPA),~\cite{12JMS-Ren} demonstrate markedly improved accuracy in reproducing hydrogen bond structures,~\cite{14JCP-DiStasio} solvation shells of hydrated water ions,~\cite{18NC-Chen} electrochemical energy levels,~\cite{16PRL-Cheng} and thermodynamic properties~\cite{19PNAS-Cheng} that are in better agreement with experimental observations.
Nevertheless, their steep computational costs severely restrict applications to large-scale systems or long-timescale dynamics, highlighting a critical challenge in balancing accuracy and efficiency for realistic modeling.

In recent years, machine learning (ML) methods utilized in electronic structure calculations emerge as a promising avenue to achieve both high accuracy and efficiency.~\cite{20JPCL-Dral,22ES-Kulik,23SCI-Huang,24AM-Aldossary}
For example, taking the electron density as the input of deep neural network, direct outputs can be designed, such as the total energy~\cite{19PRA-Ryczko,20NC-Bogojeski} and
the potential.~\cite{96JCP-Tozer,18JCP-Nagai,20SR-Ryabov}
From the data-driven perspective, methods like the contrastive learning is demonstrated to be an efficient method to learn physical constraint in the exchange energies.~\cite{23DD-Gong}
In addition, by using the atomic configurations and electronic structures as input of neural network, ML models can yield accurate electron density~\cite{23PRB-Lv} and density of states.~\cite{19npjCM-Chandrasekaran,22PRB-Zeng}

An alternative way to use ML techniques is to bypass direct outcomes but focus on designing XC functional with the aid of ML algorithms, such as prediction of the grid-based XC energy via ML algorithms.~\cite{19JPCL-Schmidt,20npjCM-Nagai,22PRR-Nagai,22JCTC-Bystrom,24JCTC-Bystrom}
For instance, Nagai {\it et al.}~\cite{22PRR-Nagai} adopted several physical asymptotic constraints to design exchange and correlation functionals separately.
Recently, the CIDER formalism proposed by Bystrom {\it et al.}~\cite{22JCTC-Bystrom}  utilizes a collection of descriptors that ensure the exchange energy adheres to the uniform scaling rule.~\cite{24JCTC-Bystrom,24PRB-Bystrom}
In addition, the XC functional can be trained effectively when the procedure of solving the KS equation is made differentiable with the usage of the automatic differentiation technique.~\cite{21PRL-Li,21PRB-Dick,21PRL-Kasim}
Another category is the $\Delta$-learning approach,~\cite{15JCTC-Ramakrishnan,20NC-Bogojeski,23JCTC-Huang,24npjCM-Jinnouchi} which learns correction terms from a lower-level base XC functional to a higher-level target XC functional.

Taking the advantages of local atomic orbital basis set in electronic structure calculations, several types of ML-algorithms can be designed to enhance the accuracy or efficiency of the traditional DFT calculations.~\cite{19JCP-Dick,19JCP-Willatt,19CS-Fabrizio,20NC-Dick,20JCP-Gastegger,21Sci-Kirkpatrick,21JCTC-Chen,22JPCA-Li,23JPCC-Ou} 
For example, several groups have developed state-of-the-art schemes that utilize a deep neural network to predict the Hamiltoinan matrix expanded via localized basis set.~\cite{19NC-Schutt,22NCS-Li,23NC-Gong,23NCS-Li,24NCS-Gong,24PRL-LiY,24PRL-LiH,24NC-Tang,24SB-Wang,24NC-Gu,23npjCM-Zhong} 
In addition, atom-centered descriptors for electronic structure can be designed based on charge density to learn the energy and force differences between a base method and a target method.~\cite{19JCP-Dick,20NC-Dick}
The DeePKS (Deep Kohn-Sham) method~\cite{21JCTC-Chen,22CPC-Chen,22JPCA-Li} projects Kohn-Sham wave functions of given systems onto a set of atomic orbitals.
Next, an atom-centered 
descriptor that remains invariant under rotation, translation, and permutation operations is designed as input of neural network.
In this regard, the exchange-correlation energy and potential can be predicted and self-consistent calculations can be performed to reach the ground state within the framework of Kohn-Sham DFT.
For practical usage, the DeePKS method can serve as a bridge between expensive quantum mechanical models and machine learning potentials.~\cite{22JPCA-Li}
To date, the DeePKS method has been applied to molecules,~\cite{21JCTC-Chen}, liquid water~\cite{22JPCA-Li} and its solvated molecules,~\cite{24JCIM-Zhang} halide perovskites,~\cite{23JPCC-Ou} etc.
%

%

%
%

%
%

The current DeePKS method falls short in accurately predicting electronic structure properties, including the Hamiltonian and its eigenvalues. 
In this study, we introduce DeePKS-ES (electronic structure), an enhanced method that aims to more accurately predict electronic structure properties, such as the Hamiltonian, as well as energy levels and the density of states.
To achieve this, we incorporate the Hamiltonian of reference systems into the loss function, in addition to energy and force labels. 
We conduct comprehensive testing of DeePKS-ES on water molecules and liquid water systems.
In particular, we have developed a universal DeePKS-ES model trained on a diverse dataset from various water systems. Our findings indicate that the universal DeePKS-ES model is both effective and precise in predicting properties of water systems.

The paper is organized as follows. In Section 2, we outline the formulas for training and applying the DeePKS-ES method. Section 3 presents the results and provides discussions. Conclusions are summarized in Section 4.


\section{Methods} 

\subsection{Kohn-Sham Equation}

When employing non-orthogonal numerical atomic orbitals (NAO)~\cite{10JPCM-Chen} as the basis set and considering only the Gamma point in the Brillouin zone under periodic boundary conditions, the Hamiltonian matrix $H$ of a given system can be written as 
\begin{equation}
H_{\mu\nu}=\langle\phi_\mu\vert T+V_\mathrm{ext}+V_{\mathrm{Hxc}}\vert\phi_\nu\rangle,
\end{equation}
where $\phi_{\mu}$ and $\phi_{\nu}$ are numerical atomic orbitals. The electron kinetic energy term, the external potential, and the Hartree and exchange-correlation terms are labeled as $T$, $V_\mathrm{ext}$, $V_\mathrm{Hxc}$, respectively.
In this regard, the Kohn-Sham equation turns into a generalized eigenvalue problem as
\begin{align}
\label{eq:HC=ESC}
H\mathbf{c}_{i}=\varepsilon_{i}S\mathbf{c}_{i},
\end{align}
where $\epsilon_i$ and $\mathbf{c}_i$ being the eigenvalue and eigenvector of electronic state $i$, respectively.
The overlap matrix is
\begin{equation}
S_{\mu\nu}=\langle\phi_\mu\vert\phi_\nu\rangle.
\end{equation}
Once the $\mathbf{c}_i$ has been solved, the real space electronic wave function can be calculated as
\begin{align}
\label{eq:wavefunction}
\psi_{i}(\mathbf{r})=\sum_{\mu}c_{i\mu}\phi_\mu(\mathbf{r}).
\end{align}
Here, $c_{i\mu}$ is the $\mu\text{-th}$ element of $\mathbf{c}_i$, representing the combination coefficient of the atomic orbital $\phi_\mu(\mathbf{r})$ in the wave function $\psi_{i}(\mathbf{r})$.
The density matrix takes the form of
\begin{align}
\rho_{\mu\nu}=\sum_{i}f_i c_{i\mu}c_{i\nu}^{\ast},
\end{align}
where $f_i$ is the occupation number.
In this regard, the real space electron density can be evaluated as
\begin{align}
\rho(\mathbf{r})=\sum_{\mu\nu}\rho_{\mu\nu} \phi_{\mu}(\mathbf{r})\phi_{\nu}(\mathbf{r}).
\end{align}
With converged electron density, the total energy of the system can be expressed as
\begin{equation}
\begin{aligned}
  E^\mathrm{tot} = & \sum_{i=1}f_i\epsilon_i - \int V_\mathrm{Hxc}(\mathbf{r})\rho(\mathbf{r}) \mathrm{d}\mathbf{r}+E_\mathrm{Hxc}[\rho(\mathbf{r})]+E_{\rm II},
\end{aligned}
\label{eq:trace_energy}
\end{equation}
where $E_\mathrm{Hxc}$ is the Hartree and exchange-correlation energy, and $E_{\rm II}$ is the ion-ion repulsive energy.

\subsection{DeePKS Model}
\label{Sec::DeePKS}

Given a target Hamiltonian $H^{\mathrm t}$ expanded with atomic orbitals and is obtained from an accurate electronic structure method, the DeePKS method adopts a density-dependent neural network to represent the difference $V^{\delta}$ between the target Hamiltonian and a base Hamiltonian $H^{\mathrm b}$.
Consequently, the total neural-network-based Hamiltonian $H^{\mathrm d}$ given by the DeePKS method can be written as
\begin{equation}
H^{\mathrm d}=H^{\mathrm b}+V^{\delta}.
\end{equation}
The $V^{\delta}$ term has a corresponding energy term $E^{\delta}$, which is designed to be the summation of contributions from individual atoms
\begin{align}
E^\delta=\sum_{I}F_{\mathrm{NN}}(\mathbf{d}^I).
\end{align}
Here, $F_{\mathrm{NN}}$ represents a neural network characterized by a set of parameters $\omega$. 
The input to this network is the descriptor $\mathbf{d}^I$, which contains electronic structure information of atom $I$.
Then, the potential $V^{\delta}$ is obtained from
\begin{equation}\begin{aligned}
V^\delta_{\mu\nu}&=\frac{\partial E^{\delta}}{\partial\rho_{\mu\nu}} \\
&=\sum_{Inlm}\frac{\partial E^{\delta}}{\partial\mathbf{d}^I_{nlm}}\frac{\partial\mathbf{d}^I_{nlm}}{\partial\rho_{\mu\nu}}.
\label{eq:v_delta}
\end{aligned}\end{equation}
Here we employ the chain rule to evaluate $V^\delta_{\mu\nu}$ by utilizing the descriptor $\mathbf{d}^I_{nlm}$, which is defined as part of $\mathbf{d}^I$, characterized by atom $I$ as well as the principal quantum number $n$, the angular momentum quantum number $l$, and the magnetic quantum number $m$.

In order to evaluate the descriptor $\mathbf{d}^I_{nlm}$ for each atom, a projected local density matrix is defined as
\begin{equation}\begin{aligned}
D_{{nlmm}^{\prime}}^I&=
\sum_{\mu\nu}\rho_{\mu\nu}\langle\phi_{\mu}|\alpha_{{nlm}}^I\rangle\langle\alpha_{{nlm}^{\prime}}^I|\phi_{\nu}\rangle.
\end{aligned}\end{equation}
Here, a set of localized orbitals $|\alpha_{nlm}^{I}\rangle$ centers at atom $I$ are needed.  
In practice, a set of GTO (Gaussian type orbital) functions and spherical Bessel functions were chosen to be $|\alpha_{nlm}^{I}\rangle$ 
when implementing the DeePKS method in PySCF~\cite{21JCTC-Chen} and ABACUS,~\cite{22JPCA-Li} respectively. Note that atoms with different species share the same set of atomic orbitals $|\alpha_{nlm}^{I}\rangle$, forming a nearly complete basis set.
Then, to preserve the rotational symmetry, a series of block matrices with the same indices of $I$, $n$, and $l$ are selected from $D_{{nlmm}^{\prime}}^I$ and diagonalized.
The resulting eigenvalues $\mathbf{d}^I_{nlm}$ are extracted and organized into vectors. In this regard, we define $\mathbf{d}^I$ as a vector containing the eigenvalues sharing the same atomic index $I$.

\begin{figure}
  \includegraphics[width=6cm]{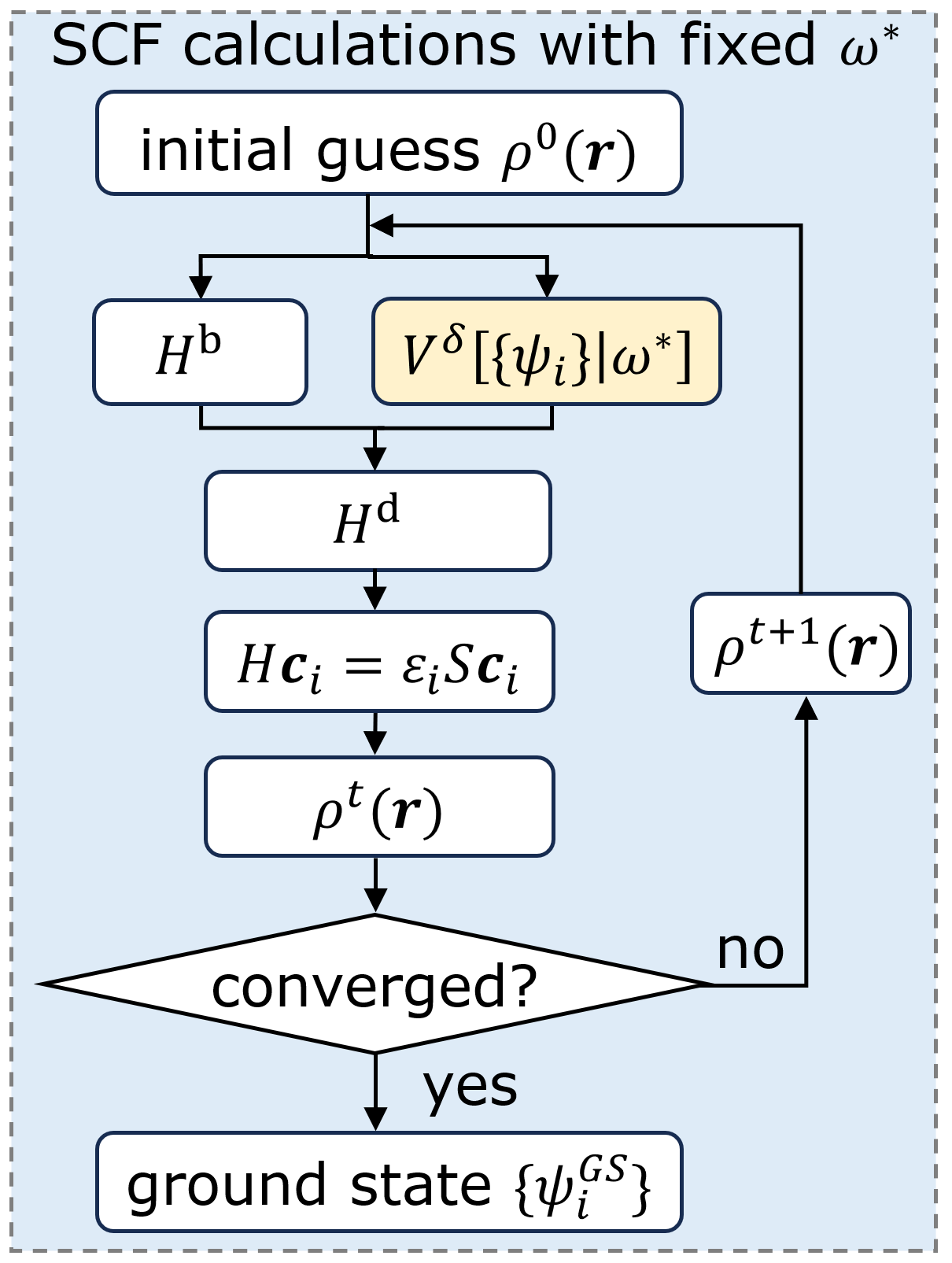}
  \caption{(Color online) 
  Workflow of the self-consistent field (SCF) calculations to solve the Kohn-Sham equation $H\mathbf{c}_i=\epsilon_{i}S\mathbf{c}_i$ with a DeePKS model.
  The Hamiltonian matrix, overlap matrix, eigenvalues, and eigenvectors are $H$, $S$, $\epsilon_i$, and $\mathbf{c}_i$, respectively. 
  The initial charge density is $\rho^0(\mathbf{r})$ and the charge density at step $t$ is $\rho^t(\mathbf{r})$.
  The total Hamiltonian matrix $H^{\rm d}$ of a given system with the DeePKS model is constructed by adding the DeePKS potential $V^\delta$ to the base Hamiltonian matrix $H^{\rm b}$.
  In the SCF calculations, $V^{\delta}$ has fixed parameters $\omega^{\ast}$ and is determined by electronic wave functions $\{\psi_i\}$.
  The ground state (GS) is attained when the charge density satisfies the convergence criterion defined in Eq.~\ref{eq:scf_thr}.
  Otherwise, the charge density is updated for the next iteration.
  }
  \label{workflow_scf}
\end{figure}

With the above definitions, $\frac{\partial E^{\delta}}{\partial\mathbf{d}^I_{nlm}}$, the first term in the last expression of Eq.~\ref{eq:v_delta} for each atom $I$, can be evaluated via back-propagation defined in the neural network framework.
While the expression for the second term can be further decomposed as
\begin{equation}\begin{aligned}
    \frac{\partial \mathbf{d}^I_{nlm}}{\partial\rho_{\mu\nu}}=&\sum_{m^{\prime}m^{\prime\prime}}\frac{\partial \mathbf{d}^I_{nlm}}{\partial D^I_{nlm^{\prime}m^{\prime\prime}}}
    \frac{\partial D^I_{nlm^{\prime}m^{\prime\prime}}}{\partial \rho_{\mu\nu}}
    \\
    =&\sum_{m^{\prime}m^{\prime\prime}} \frac{\partial \mathbf{d}^I_{nlm}}{\partial D^I_{nlm^{\prime}m^{\prime\prime}}}\langle\phi_\mu\vert\alpha^I_{nlm^{\prime}}\rangle\langle\alpha^I_{nlm^{\prime\prime}}\vert\phi_\nu\rangle.
\label{eq:d-rho}
\end{aligned}\end{equation}
The first term of the above equation can be again evaluated with the backpropogation algorithm, even without a neural network.
The second term can be calculated with the two given sets of atomic orbitals.
Finally, the correction potential $V^{\delta}_{\mu\nu}$ can be obtained to perform self-consistent calculations, as illustrated in Fig.~\ref{workflow_scf}.

After the self-consistent iteration reaches convergence, the total energy with the DeePKS model is calculated as
\begin{equation}
\begin{aligned}
  E^\mathrm{d} = &\sum_{i=1}f_i\epsilon_i - \int V_\mathrm{Hxc}(\mathbf{r})\rho(\mathbf{r}) \mathrm{d}\mathbf{r} - \sum_{\mu\nu}\rho_{\mu\nu}V_{\mu\nu}^{\delta}\\&+E_\mathrm{Hxc}[\rho(\mathbf{r})]+ E^\delta + E_\mathrm{II}.    
\end{aligned}
\end{equation}

The force acting on the atom $I$ can be computed from
\begin{equation}
\mathbf{F}^\mathrm{d}_{I}=-\frac{\partial E^\mathrm{d}}{\partial \tau_{I}},
\end{equation}
where $\tau_I$ is the position of atom $I$. 
Note that when utilizing numerical atomic orbitals as the basis set, in addition to the force contribution from the Ewald term, three additional force terms require evaluation.
Namely, the Feynman-Hellmann force, the Pulay force, and the force that arises from the nonorthogonality of the atomic orbitals. In fact, the presence of the $E^{\delta}$ term contributes to all of the three forces except for the Ewald term.

First, consider the derivative taken when the density matrix $\rho_{\mu\nu}$ stays constant.
It contains two components, one is the Feynman-Hellmann force coming from the Hamiltonian while the other one is the Pulay force coming from the movement of atomic orbitals.
The correction for these two terms from DeePKS models is 
\begin{equation}\begin{aligned}
\mathbf{F}^{\delta}_{I}& =-\frac{\partial E^{\delta}}{\partial\tau_I}
=-\sum_{Inlm}\frac{\partial E^{\delta}}{\partial\mathbf{d}^I_{nlm}}\frac{\partial\mathbf{d}^I_{nlm}}{\partial\tau_I} \\
&=-\sum_{Inlm}\frac{\partial E^\delta}{\partial \mathbf{d}^I_{nlm}}\sum_{m^{\prime}m^{\prime\prime}}\frac{\partial \mathbf{d}^I_{nlm}}{\partial D^I_{nlm^{\prime}m^{\prime\prime}}} \frac{\partial D^I_{nlm^{\prime}m^{\prime\prime}}}{\partial\tau_{I}} ,
\end{aligned}\end{equation}

Second, consider the derivative due to the change in the density matrix $\rho_{\mu\nu}$ with atomic movements.
It arises from the fact that NAOs are not orthogonal, and the contribution is evaluated via
\begin{equation}
\mathbf{F}^{\mathrm{ortho}}_{I}=-\sum_{\mu\nu}E_{\mu\nu}\frac{\partial S_{\mu\nu}}{\partial \tau_I},
\end{equation}
where $E_{\mu\nu}$ is the energy density matrix that takes the form
\begin{equation}
E_{\mu\nu}=\sum_{i}f_i\epsilon_ic_{i\mu}c_{i\nu}^{\ast}.
\end{equation}
Note that the eigenvalue $\epsilon_i$ contains information from the $V^{\delta}$ term. More information about the force terms can be seen in Refs.~\citenum{02JPCM-Soler,24WCMS-Lin}.
In order to demonstrate the correct implementation of atomic forces within the DeePKS scheme, a finite difference test was carried out and the results are shown in Supplementary Section \MakeUppercase{\romannumeral 1}.

\subsection{Iterative Training}

\begin{figure*}[htbp]
  \includegraphics[width=16cm]{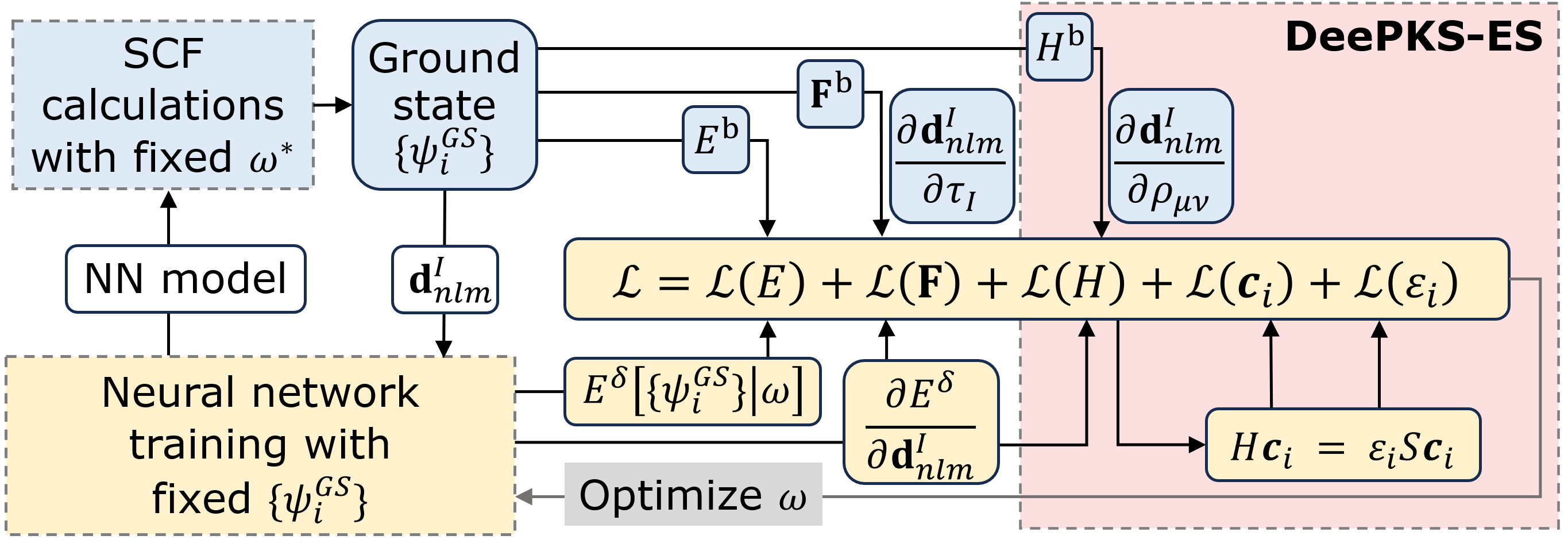}
  \caption{(Color online)
  Workflow for training a DeePKS-ES model consists of two iterative steps. 
  The first step is conducting self-consistent field (SCF) calculations with fixed parameters $\omega^{\ast}$ of the neural network (NN) model to yield ground sttate wave function $\{\psi_{i}^{GS}\}$.
  The second step is training NN with fixed $\{\psi_{i}^{GS}\}$, implying the input descriptors $\mathbf{d}^{I}_{nlm}$ are also fixed.
  The model parameters $\omega$ are optimized according to the loss function $\mathcal{L}$, consisting of $\mathcal{L}(E)$, $\mathcal{L}(\mathbf{F})$, $\mathcal{L}(H)$, $\mathcal{L}(\mathbf{c_i})$, $\mathcal{L}(\epsilon_i)$, which correspond to loss functions of total energy, atomic forces, Hamiltonian matrix, wave function coefficients, and energy levels, respectively.
  The latter three terms included in the red box are newly added in the DeePKS-ES model.
  The loss function represents the difference between the properties provided by the DeePKS-ES model and those of the target functional, as shown in Eq.~\ref{eq:lossfn}.
  The total energy of the DeePKS-ES model is obtained by adding the NN output $E^{\delta}[\{\psi_{i}^{GS}\}|\omega]$ to the total energy from the base functional $E^{\rm b}$.
  For atomic forces, both $\frac{\partial E^{\delta}}{\partial\mathbf{d}^I_{nlm}}$ provided by the NN model and an additional term $\frac{\partial\mathbf{d}_{nlm}^{I}}{\partial\tau_{I}}$ obtained after SCF calculations are used to construct $\mathbf{F}^{\delta}$, which is subsequently added to $\mathbf{F}^{\rm b}$.
  Similarly, the Hamiltonian matrix comprises $H^{\rm b}$ and $H^{\delta}$, with the latter constructed by $\frac{\partial E^{\delta}}{\partial\mathbf{d}^I_{nlm}}$ and $\frac{\partial\mathbf{d}_{nlm}^{I}}{\partial\rho_{\mu\nu}}$.
  Here $\tau_I$ is the atomic position of atom $I$ while $\rho_{\mu\nu}$ is the density matrix.
  After the Hamiltonian of the DeePKS-ES model $H$ is obtained, the Kohn Sham equation $H\mathbf{c}_i=\epsilon_{i}S\mathbf{c}_i$ with $S$ being overlap matrix is solved to obtain eigenvectors $\mathbf{c}_i$ and eigenvalues $\epsilon_i$.
  }\label{workflow}
\end{figure*}

Fig.~\ref{workflow} illustrates the procedures of training a DeePKS model. Subject to the orthogonal condition of electronic wave functions $\langle\psi_i|\psi_j\rangle=\delta_{ij}$, 
the training of a DeePKS model is to optimize the following formula
\begin{equation}
\begin{aligned}
\label{eq:two-min}
&\min_{\boldsymbol{\omega}}\mathbb{E}_{\text{data}}\Big[\Big(E^{\rm t}-\min_{\{\psi_i\}}E^{\rm d}[\{\psi_i\}|\omega]\Big)^2\Big].
\end{aligned}
\end{equation}
Here, $\mathbb{E}_{\text{data}}$ represents the average over the training data.
The total energy with the DeePKS model $E^{\rm d}$ is decided by both the wave functions ${\{\psi_i\}}$ that determine the electron density and $\mathbf{d}^I_{nlm}$, and the DeePKS model parameters $\omega$ that determine the XC functional.
To optimize the parameters $\omega$, the energy difference between the target total energy $E^{\rm t}$ and the total energy $E^{\rm d}$ obtained after an SCF loop is defined as part of the loss function.

The overall training process for a DeePKS model comprises a few iterations until convergence is achieved. 
For each iteration, Fig.~\ref{workflow} illustrates a two-step scheme to optimize the DeePKS model.
In the first step, as shown in the blue box, with the parameters $\omega$ of the DeePKS neural network (NN) model fixed as $\omega^*$, the Kohn-Sham equation is solved by SCF calculations performed with ABACUS~\cite{10JPCM-Chen,16CMS-Li} to obtain the ground state information. 
Then, a set of descriptors $\{\mathbf{d}^{I}\}$ for each atom are generated.
In the second step, as shown in the yellow box, with the wave functions fixed as $\{\psi_i^{GS}\}$, which means the input descriptors fixed as $\mathbf{d}^{I}_{nlm}$, the model parameters $\omega$ are updated over multiple epochs using back-propagation algorithms provided by the DeePKS-kit.~\cite{22CPC-Chen}
Typically, $\omega$ are set to zero at the beginning, implying that the first wave functions are actually the ground state wave functions of the base model.

%
%

\subsection{Loss Function}

During the training procedure for the DeePKS model, 
the loss function can be treated as minimizing the loss function 
\begin{equation}
\mathcal{L}=\mathbb{E}_\text{data}{\Big[(E^{\rm t}-E^{\rm d})^2\Big]},
\end{equation}
For the sake of clarity and brevity, we omit the explicit notation of $\{\psi_i\}$ and $\omega$.
The above equation only contains the total energy difference, additional properties such as atomic forces, stress, and band gap can be added to the loss function.~\cite{22JPCA-Li,23JPCC-Ou}

We propose a revised DeePKS method named DeePKS-ES, the key procedure of which is shown in the red box of Fig.~\ref{workflow}. The loss function in the DeePKS-ES method is defined as
\begin{equation}\begin{aligned}
\label{eq:lossfn}
\mathcal{L}&=\mathbb{E}_{\text{data}}\Big[\lambda_{\rm E}\Big(\frac{E^{\mathrm t}-E^{\mathrm d}}{N_{\rm a}}\Big)^2 \\
&+\lambda_{\rm F}\frac{\sum_{I=1}^{N_{\rm a}}\Vert\mathbf{F}_{I}^{\mathrm t}-\mathbf{F}_{I}^{\mathrm d}\Vert^2}{3N_{\rm a}} \\
&+\lambda_{\rm H}\frac{\Vert H^{\mathrm t}-H^{\mathrm d}\Vert^2}{N_{l}} \\
&+\lambda_{\psi}\frac{\sum_{i=1}^{N_{\rm \psi}}\Vert\mathbf{c}_{i}^{\mathrm t}-\mathbf{c}_{i}^{\mathrm d}\Vert^2}{N_{l}{N_{\rm \psi}}} \\
&+\lambda_{\rm b}\frac{\sum_{i=1}^{N_{\rm b}}(\varepsilon_{i}^{\mathrm t}-\varepsilon_{i}^{\mathrm d})^2}{N_{\rm b}}\Big],
\end{aligned}\end{equation}
where five terms are displayed, and the expression $\Vert x\Vert^2$ denotes the square and sum of each element of $x$.
The parameters $\lambda_{\rm E}, \lambda_{\rm F}, \lambda_{\rm H}, \lambda_{\rm \psi}, \lambda_{\rm b}$ denote weighting factors of the energy, atomic forces, Hamiltonian matrix, wave function coefficients, energy levels, respectively.
The superscripts $\mathrm t$ and $\mathrm d$ represent the results given by the target functional and the DeePKS-ES model, respectively.
The first term is the energy term defined as the average energy difference per atom compared to the target energy, with $N_{a}$ being the number of atoms in the system.
The second term involves the atomic forces, the force differences along the three directions for each atom are evaluated.

In addition, the Hamiltonian matrix represented with the localized basis set, as well as the eigenvectors and eigenvalues, are added to the loss function as the third, fourth, and fifth terms, respectively.
In the Hamiltonian term, the difference between $H^{\mathrm t}$ and the Hamiltonian matrix $H^{\mathrm d}$ obtained from the DeePKS-ES model is evaluated. Specifically, the differences of each matrix element are squared and summed. 
%
We note that the size of the Hamiltonian matrix is $N_{l} \times N_{l}$, with $N_{l}$ denoting the number of localized basis set.
However, the number of neighbors for each atom does not increase with system size.
This suggests that as the system size increases, the number of nonzero matrix elements in the Hamiltonian matrix increases proportionally to $O(N_{l})$ rather than $O(N_{l}^2)$. 
Therefore, we divide the sum by $N_{l}$ instead of $N_{l}^2$.
The loss term involving eigenvectors includes $N_{\rm \psi}$ wave functions, each represented by the coefficient vector $\mathbf{c}_i$, which contains $N_{l}$ elements. 
The final term in the loss function pertains to the eigenvalues, which represent the mean square of the difference of $N_{\rm b}$ energy levels.
The two factors $N_{\rm \psi}$ and $N_{\rm b}$ refer to the needed number of eigenvectors $\{\mathbf{c}_i\}$ and eigenvalues $\{\varepsilon_{i}\}$ included in the loss function.

%

In order to train the DeePKS-ES model, we need to evaluate $H^{\mathrm d}$, which is composed of $H^{\mathrm b}$ contributed from the base functional and $V^\delta$ from the neural network adopted in the DeePKS-ES model. As shown in Eq.~\eqref{eq:v_delta}, the first term of $V^\delta$ is $\frac{\partial E^{\delta}}{\partial \mathbf{d}_{nlm}^{I}}$, which is evaluated from the back-propagation algorithm implemented in the DeePKS-kit package.
The second term $\frac{\partial\mathbf{d}^I_{nlm}}{\partial\rho_{\mu\nu}}$ is independent of the model parameters $\omega$ and can be calculated via Eq.~\ref{eq:d-rho}.
This term is obtained following the SCF calculations executed in ABACUS.
In practice, the two terms in Eq.~\ref{eq:d-rho}, namely, $\frac{\partial \mathbf{d}^I_{nlm}}{\partial D^I_{nlm^{\prime}m^{\prime\prime}}}$
and $\langle\phi_\mu\vert\alpha^I_{nlm^{\prime}}\rangle$, are output separately. 
The last two terms of the loss function Eq.~\ref{eq:lossfn} involve the eigenvectors and eigenvalues of the Hamiltonian matrix, which can be derived from the diagonalization of the target Hamiltonian and the DeePKS-ES Hamiltonian matrix.

\subsection{Model training and evaluation}

In this study, we developed and evaluated two types of DeePKS-ES models, each based on distinct training data sets.
Initially, we developed system-specific DeePKS-ES models, referred to as DeePKS-ES(s), by training them solely on individual systems, including water monomers, dimers, hexamers, and the liquid water phase.
Subsequently, we constructed a comprehensive model for aqueous systems, termed DeePKS-ES(u), which was trained on a combined dataset encompassing various water molecule configurations as well as the liquid state.
In both models, we chose the PBE exchange-correlation functional as the base model and the HSE06 functional,~\cite{03JCP-Jochen,06JCP-Krukau} hereafter referred to as HSE, as the target model.

When training the DeePKS-ES models, we chose the neural network as a fully-connected neural network with 3 hidden layers, each hidden layer has 100 neurons. 
For the loss function, we set $\lambda_{\rm E}, \lambda_{\rm F}, \lambda_{\rm H}, \lambda_{\rm \psi}, \lambda_{\rm b}$ as 1, 1, 0.005, 0.1, 1, respectively. The parameter $N_{\rm \psi}$ is set to the number of occupied states $N_{\rm occ}$, and $N_{\rm b}$ is set to $2n_{\rm occ}$ to include some unoccupied states.
The number of training iterations is 9, 8, 5, 8, and 7 for the water monomer, water dimer, water hexamer, 32-molecule liquid water, and the universal training models, respectively.

Finally, we define root mean square errors (RMSEs) to evaluate the accuracy of different models. The five types of RMSEs for total energy, atomic forces, hamiltonian matrix, eigenvectors, and eigenvalues are respectively defined as follows
\begin{align}
\label{eq:RMSE-begin}
\Delta E &= \frac{\lvert E^{\rm t}-E^{\rm d} \rvert}{N_{\rm a}} \\
\Delta \mathbf{F} &= \sqrt{\frac{\sum_{I=1}^{N_{\rm a}}\Vert\mathbf{F}_{I}^{\rm t}-\mathbf{F}_{I}^{\rm d}\Vert^2}{3N_{\rm a}}} \\
\Delta H &= \sqrt{\frac{\Vert H^\text{t}-H^{\text{d}}\Vert^2}{N_{l}}} \\
\Delta \mathbf{c} &= \sqrt{\frac{\sum_{i=1}^{N_{\rm occ}}\Vert\mathbf{c}_{i}^{\rm t}-\mathbf{c}_{i}^{\rm d}\Vert^2}{N_{l}{N_{\rm occ}}}} \\
\label{eq:RMSE-end}
\Delta \varepsilon &= \sqrt{\frac{\sum_{i=1}^{N_{\rm occ}+1}(\varepsilon_{i}^{\rm t}-\varepsilon_{i}^{\rm d})^2}{N_{\rm occ}+1}},
\end{align}
where $N_{\rm occ}$ is the number of occupied states.
Note that the RMSE here is similar to those defined in Eq.~\eqref{eq:lossfn}. 
The main difference is that the energy level defined here considers the $N_{\rm occ}+1$ energy levels.


\subsection{Dataset Preparation}

The training dataset of the water monomer, water dimer, water hexamer, 32-molecule liquid water, and the universal models contained 750, 300, 380, 300, and 275 configurations, respectively.
Specifically, the 275 configurations in the universal model contained 25, 50, 50, 150 sets of data for the water monomer, water dimer, water hexamer, and 32-molecule liquid water systems, respectively. 
In water molecular systems, relatively large cell lengths, i.e. 28, 37.8, and 37.8 Bohr were chosen to simulate water monomer, water dimer, and water hexamer molecules in a vacuum, respectively.
In the dataset of water hexamer clusters, four stable configurations were included, i.e., prism, cage, book, and cyclic structures.
To maintain the density of liquid water in periodic systems with 32 and 64 water molecules, cell lengths of 18.6 and 23.5 Bohr were selected, respectively.
All of the above configurations were generated by first-principles molecular dynamics simulations. Both molecular dynamics and DFT calculations were performed by the ABACUS\cite{10JPCM-Chen,16CMS-Li} package with periodic boundary conditions.
A kinetic energy cutoff of 100 Ry was applied, together with the norm-conserving Vanderbilt (ONCV) pseudopotentials~\cite{15CPC-Schlipf} generated with the PBE exchange-correlation functional.
In SCF iterations, the convergence criterion for charge density was $1.0\times10^{-6}$.
Specifically, the convergence criterion was defined as 
\begin{equation}\begin{aligned}
\label{eq:scf_thr}
\frac{1}{N_e} \int \lvert \Delta \rho(\mathbf{r}) \rvert d\mathbf{r},
\end{aligned}\end{equation}
where $N_e$ is the number of electrons.
$\Delta \rho(\mathbf{r})$ is the charge density difference between two sequential SCF iterations.
The Broyden charge mixing method~\cite{88PRB-Johnson} was employed with the ratio of adding the new charge density set as 0.4.

The NAO bases were chosen as double-zeta orbitals with a polar orbital (DZP) with raidus cutoff being 10.0 Bohr.
When utilizing the DeePKS method,
the projected orbitals $|\alpha^I_{nlm}\rangle$ were chosen to be spherical Bessel functions~\cite{10JPCM-Chen} with the energy and radius cutoff being 200 Ry and 2.0 Bohr, respectively.
For the water systems tested in this work, we selected nine sets of $s$, $p$, and $d$ orbitals, resulting in a total of 81 projected orbitals.
The HSE results were calculated with the LibRI v0.1.0 package,~\cite{20JPCL-Lin} which utilizes the resolution of the identity (RI) method that substantially reduces the computational costs with NAO bases.

\section{Results}

\begin{figure}
  \includegraphics[width=8cm]{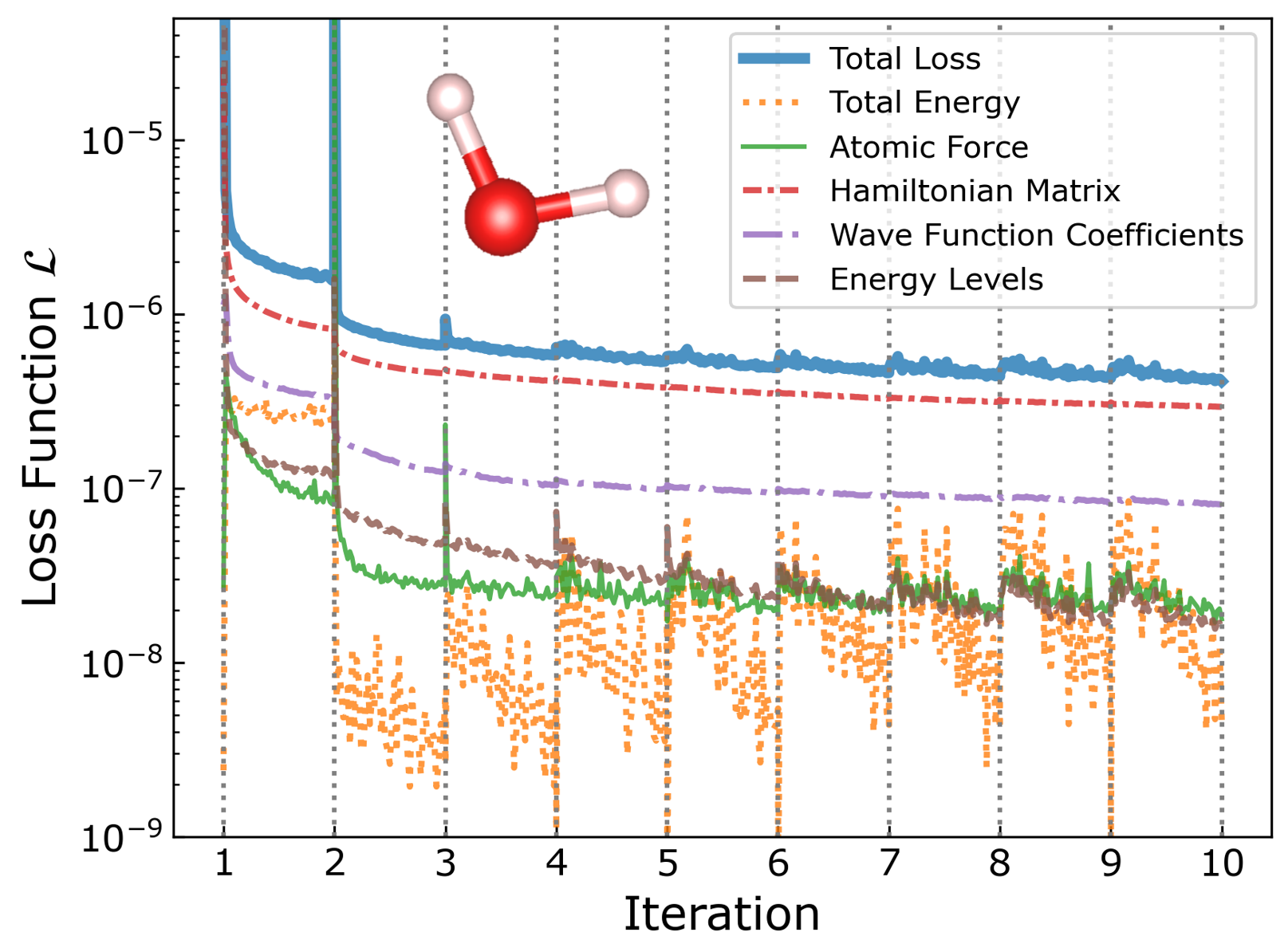}
  \caption{(Color online)
  Training loss curve of the water monomer model on a logarithmic scale. 
  Variations in the total loss function $\mathcal{L}$ (thick line on the top) and its components, including total energy, atomic forces, Hamiltonian matrix, wave function coefficients, and energy levels, are detailed.
  }\label{trn_loss}
\end{figure}

%

\subsection{Water Monomer}

We generate a DeePKS-ES(s) model for the water monomer system.
Fig.~\ref{trn_loss} shows changes of the loss function with respect to 9 iterations during the model training.
The number on the horizontal axis depicts the starting training step for each iteration, while the value of the loss function is on the vertical axis.

\begin{figure*}[htbp]
  \includegraphics[width=16cm]{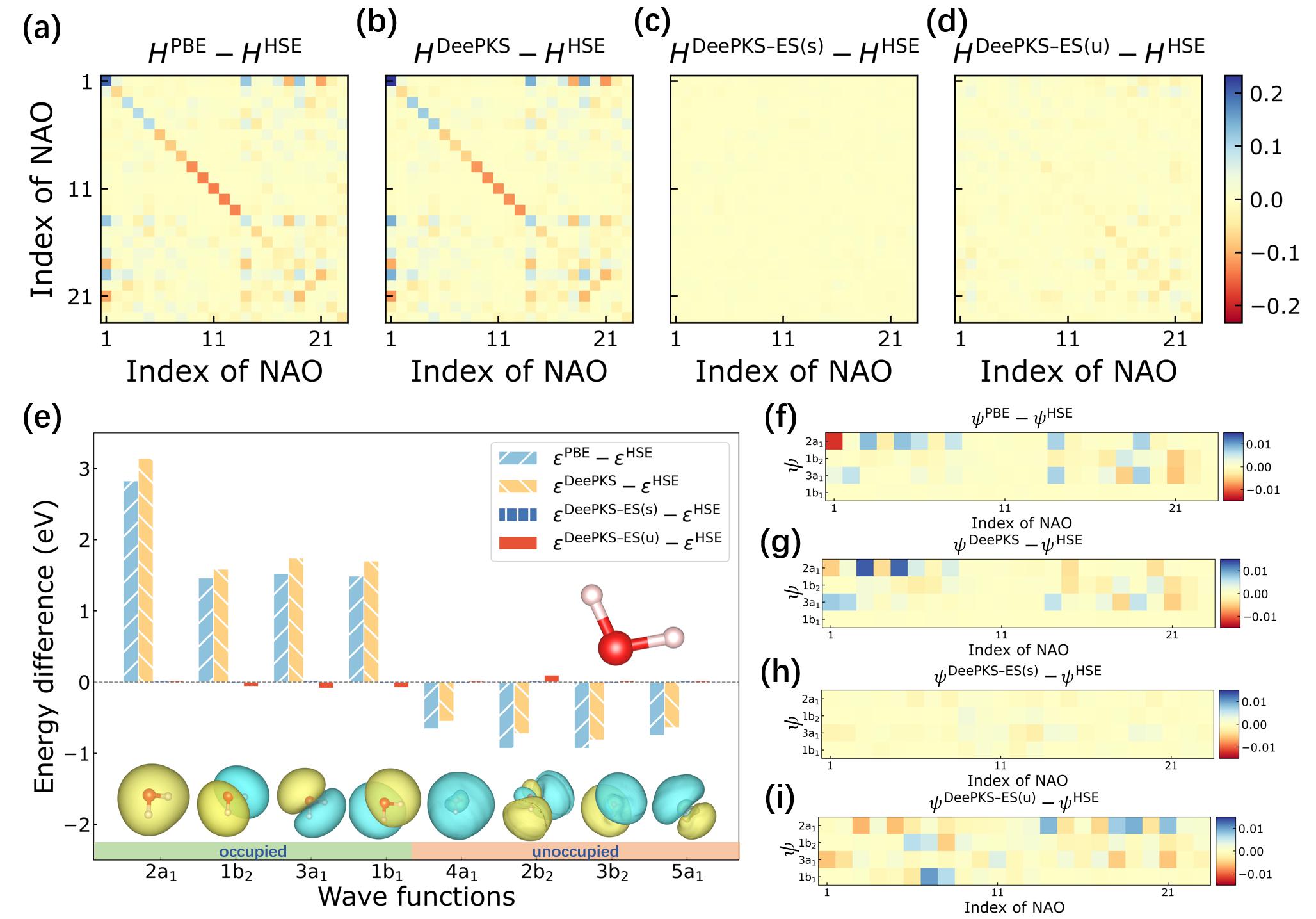}
  \caption{(Color online)
  Electronic structural property differences among the PBE, DeePKS, DeePKS-ES(s), and DeePKS-ES(u) methods, compared to the reference HSE functional, are studied within a water monomer test configuration.
  The label DeePKS-ES(s) denotes a model trained solely on data from the water monomer, whereas DeePKS-ES(u) refers to a universal approach trained on a dataset encompassing various systems.
  Figures (a-d) illustrate deviations in the Hamiltonian matrix, with the axes representing indices of the NAO (numerical atomic orbital) basis set.
  Figure (e) displays energy differences across occupied and unoccupied energy levels, alongside their respective wave functions.
  Figures (f-i) present discrepancies in the wave functions. The horizontal axis correlates to the NAO basis index, and the vertical axis corresponds to the wave function values.
  Hence, the discrepancy observed at the intersection of row $i$ and column $u$ corresponds to the difference in $\mathbf{c}_{i\mu}$ from Eq.~\ref{eq:wavefunction}.}
  \label{difference_monomer}
 \end{figure*}

We find all of the loss terms generally decrease both within the training steps and between the iterations. 
In particular, the loss of energy term decreases fast in the first two iterations. Then, the energy loss term exhibits fluctuations but keeps at the smallest values after each iteration when compared to the other four terms.
Furthermore, the loss terms of the Hamiltonian and wave function coefficients decrease smoothly, even between different iterations.
Overall, as the number of iterations increases, the total loss value converges to a small value.
In detail, the loss values for total energy, atomic forces, Hamiltonian matrix, eigenvectors, and eigenvalues are around $10^{-9}, 10^{-8}, 10^{-7}, 10^{-8}, 10^{-8}$.
As a result, the above loss values correspond to the difference of around $10^{-3}$ eV/atom, $10^{-3}$ eV/\AA, $10^{-3}$ Ry, $10^{-3}$, and $10^{-3}$ eV for the total energy, atomic forces, Hamiltonian matrix, eigenvectors, and eigenvalues.

%

We employ a water monomer that is not included in any training dataset to validate the DeePKS models.
When compared to the total energy and atomic forces obtained from HSE calculations, the original DeePKS model performs the best, with total energy difference and root mean square error (RMSE) of forces being $1.06\times10^{-5}$ eV/atom and $6.14\times10^{-3}$ eV/\AA, respectively. 
Both DeePKS-ES(s) and DeePKS-ES(u) models are optimized with more terms in the loss function and perform slightly worse than the original DeePKS model on the two properties. 
Specifically, the differences of total energy and atomic forces are $2.60\times10^{-4}$ eV/atom, $6.14\times10^{-3}$ eV/\AA~ 
for the DeePKS-ES(s) model, and $2.67\times10^{-2}$ eV/atom, $1.18\times10^{-2}$ eV/\AA~ for the DeePKS-ES(u) model.
Still, the above models perform better than PBE, with the energy and force differences as $4.60\times10^{-2}$ eV/atom and  $3.58\times10^{-1}$ eV/\AA, respectively.
The detailed values are also shown in in Supplementary Table S1.

Next, we turn to the discussion of Hamiltonian and related electronic structure properties.
We set four occupied states used in the loss function. At the level of using the DZP (double $\zeta$ functions plus one polar orbital) basis set, the total number of basis in a water monomer system is 23, with the first 13 orbitals centered on the oxygen atom while 5 orbitals for each of the two hydrogen atoms. Therefore, the size of a Hamiltonian matrix is 23$\times$23.
Fig.~\ref{difference_monomer} compares the differences in the Hamiltonian matrix, energy levels, and coefficients of wave functions among various methods, including the PBE functional, the original DeePKS model, the DeePKS-ES(s) model, the DeePKS-ES(u) model, with respect to the target functional HSE.
Fig.~\ref{difference_monomer}(a) shows the Hamiltonian matrix difference between the base matrix $H_{\rm PBE}$ and the target matrix $H_{\rm HSE}$.
We find the two matrices differ mostly on the on-site matrix elements $H_{ii}$ associated with the oxygen atom, with a difference up to 0.203 Ry.
Discrepancies also exist in the off-site elements ($H_{ij}, i\neq j$) between the orbitals of two hydrogen atoms and between the $s/p$ orbitals of the oxygen atom and those of the hydrogen atoms.
Remarkably, a comparison between Figs.~\ref{difference_monomer}(a) and (b) reveals that the original DeePKS model, trained solely on total energy and atomic forces, predicts the Hamiltonian matrix in a manner nearly identical to the PBE prediction. This suggests that the original DeePKS model captures minimal information regarding the Hamiltonian matrix.
Interestingly, by including the Hamiltonian matrix along with its eigenvalues and eigenvectors into the loss function and training new models, the results are substantially improved. 
For instance, Fig.~\ref{difference_monomer}(c) shows that the DeePKS-ES(s) model predicts a Hamiltonian matrix that is almost identical to the one from HSE, with a maximum difference of 0.013 Ry.
Furthermore, Fig.~\ref{difference_monomer}(d) illustrates that the universal model DeePKS-ES(u) predicts slightly worse than DeePKS-ES(s), with its maximum error being 0.056 Ry. However, the results are still satisfactory when compared to the original DeePKS model.
For the universal model, the main differences arise from the matrix elements associated with the interactions of two hydrogen atoms.
This can be understood because the distances between hydrogen and oxygen atoms and the related strength of hydrogen bonds vary from a single water molecule to clusters and liquid water.~\cite{17PNAS-Chen} 
In this regard, it will be more challenging for the universal model to capture all the physics than for a single model to capture properties in a single system.

Fig.~\ref{difference_monomer}(e) shows the differences of eight energy levels for a water monomer from different models.
The corresponding 8 molecular orbitals, including four occupied states, $\rm 2a_1$, $\rm 1b_2$, $\rm 3a_1$, $\rm 1b_1$, and four unoccupied states, $\rm 4a_1$, $\rm 2b_2$, $\rm 3b_2$, $\rm 5a_1$
are also depicted.
%
Detailed values for the energies obtained from various methods are shown in Supplementary Table S2.
Given that the energy levels are the eigenvalues of the Hamiltonian matrix, we find that the prediction accuracy of DeePKS-ES(s) and DeePKS-ES(u) models are also substantially better than the original DeePKS model.
For example, when compared to HSE, the PBE functional overestimates occupied energy levels and underestimates unoccupied ones, leading to an underestimation of the band gap by 2.15 eV.
Similarly, the original DeePKS model does not show improvements and still underestimates the band gap by 2.25 eV.
Impressively, the band gap errors from the DeePKS-ES(s) and DeePKS-ES(u) models are $1.67\times10^{-3}$ and $8.39\times10^{-2}$ eV, respectively.

Finally, Figs.~\ref{difference_monomer}(f-i) illustrates the differences in the coefficients of the four occupied wave functions, which are eigenvectors of the Hamiltonian matrix.
We find the original DeePKS model in Fig.~\ref{difference_monomer}(g) does not show 
improvement over the base PBE functional in Fig.~\ref{difference_monomer}(f), and the RMSEs for both models with respect to HSE are around $2.80\times10^{-3}$.
However, the results from the DeePKS-ES(s) model compare well with those from HSE, with the RMSE of $7.62\times10^{-4}$.
For the DeePKS-ES(u) model, the RMSE is $3.05\times10^{-3}$, which is similar to that of PBE.
This may be related to the fact that the information of wave function coefficients is not well captured in the current model.

\subsection{Eigenvalues from DeePKS-ES}

\begin{figure*}
  \includegraphics[width=16cm]{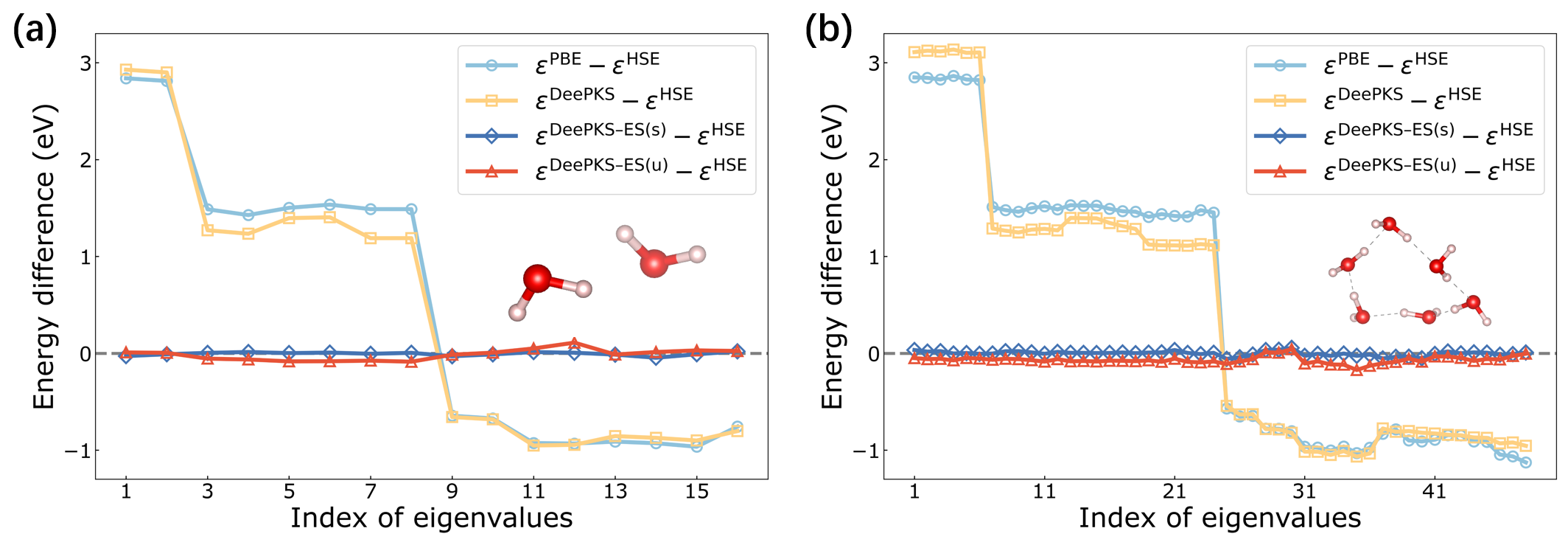}
  \caption{(Color online)
  Energy level discrepancies among the PBE, DeePKS, DeePKS-ES(s), and DeePKS-ES(u) methods relative to the reference HSE functional. 
The plotted energy levels are double the number of occupied states for given systems.
The inset depicts the molecular configuration associated with these differences.
(a) Discrepancies within a water dimer system.
(b) Discrepancies within a water hexamer system.
  }\label{difference_dimer_hexamer}
\end{figure*}

Next, we train models and test them in more systems, which include water dimers, water clusters, and liquid water. In general, 
we find the conclusions are similar to those obtained from the water monomer system. 
In particular, we focus on the eigenvalues of the Hamiltonian matrix since they are important electronic structure properties of given systems.
Figs.~\ref{difference_dimer_hexamer} (a) and (b) show the differences of 16 and 48 energy levels in the water dimer and hexamer systems, respectively.
Detailed values of the differences are shown in Supplementary Tables S3 and S4.
The structure of the water hexamer system is the cyclic structure.
We see that the energy levels from the PBE functional substantially deviate a few eV from the HSE results.
As expected, the original DeePKS model shows little improvement over the PBE functional.
Importantly, the predictions from the DeePKS-ES(s) and DeePKS-ES(u) models closely align with those energy levels from HSE, with small deviations to be around 0.01 and 0.05 eV, respectively.

\begin{figure}
  \includegraphics[width=8cm]{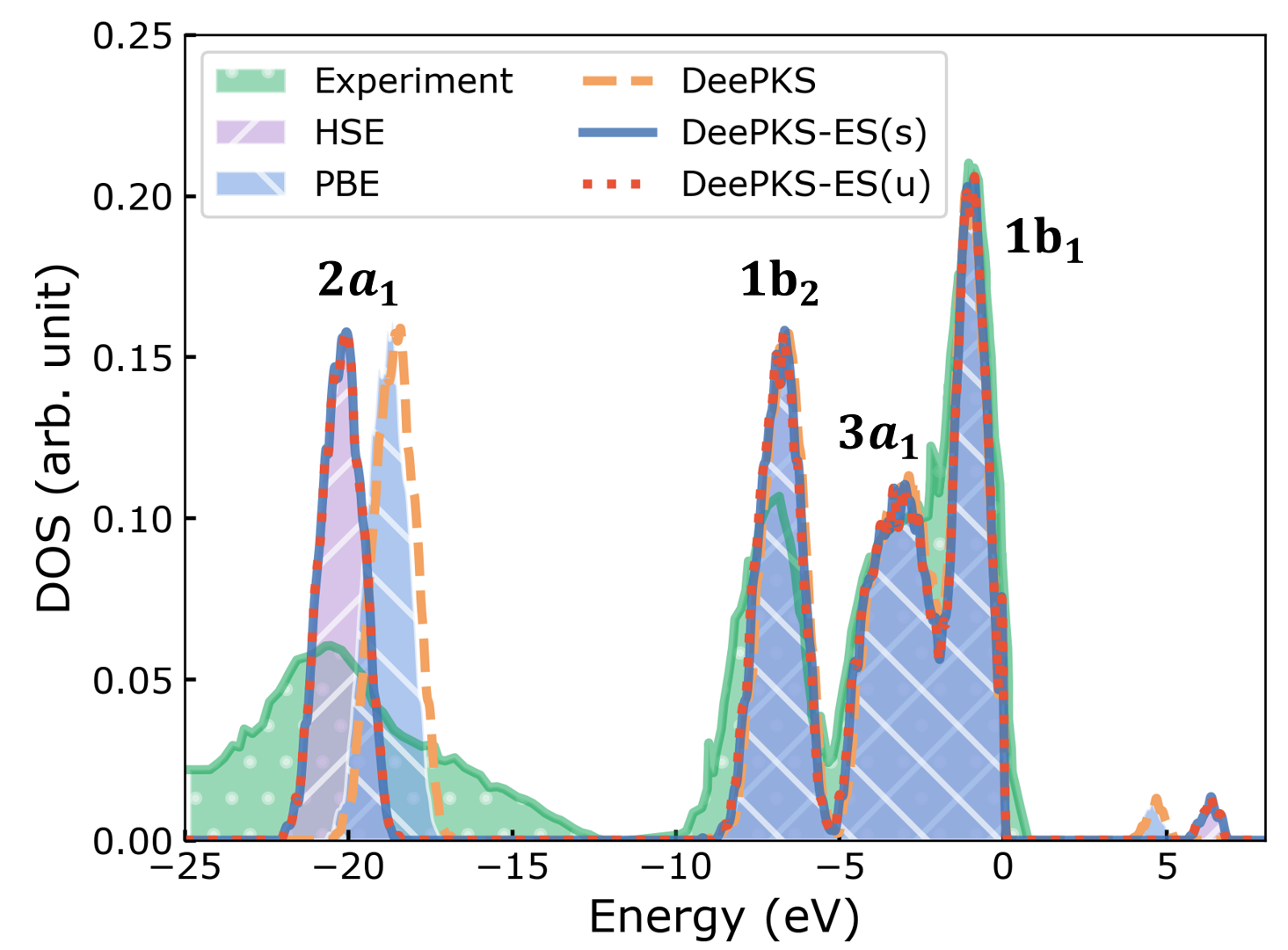}
  \caption{(Color online)
  Density of states (DOS) for a periodic system featuring 32 water molecules.
  The used methods include the HSE, PBE, DeePKS, DeePKS-ES(s), and DeePKS-ES(u), as well as the experimental data from photoemission spectroscopy~\cite{04JPCA-Winter}.
  The DOS data are aligned\cite{16PRL-Chen} to the $\mathrm{1b_1}$ peak of the experimental data.
  }\label{dos}
\end{figure}

For liquid water systems that contain 32 water molecules with periodic boundary conditions, 
Fig.~\ref{dos} shows the density of states (DOS) predicted by the HSE, PBE, DeePKS, DeePKS-ES(s) and DeePKS-ES(u) methods, as well as the experimental DOS measured by full valence band photoemission spectroscopy.~\cite{04JPCA-Winter}
Based on the spatial symmetries of the water molecule, the four peaks of the DOS are attributed to the $\mathrm{2a_1}$, $\mathrm{1b_2}$, $\mathrm{3a_1}$, and $\mathrm{1b_1}$ orbitals, consistent with the case in the water monomer systems.
The DOS for each method is averaged over 80 configurations, derived from Ab initio molecular dynamics (AIMD) simulation at 300 K.
All the simulated DOS are aligned at the position of the $\mathrm{1b_1}$ peak of the experimental data\cite{16PRL-Chen}.
%
%
Owning the highest accuracy among these computational methods, the HSE functional predicts the peak position of $\mathrm{2a_1}$ state 0.5 eV higher than the experimental data, which is substantially better than the 1.8 eV by using the PBE functional.
Not surprisingly, the original DeePKS model predicts almost the same position of the $\mathrm{2a_1}$ peak as PBE.
Again, both DeePKS-ES(s) and DeePKS-ES(u) models show excellent agreement with the results from the HSE functional, with the peak position difference being less than 0.05 eV.
Additionally, the band gap of liquid water can be determined by subtracting the energy of the first unoccupied level from that of the highest occupied level. 
The HSE functional predicts the band gap to be 7.6 eV, which is close to the experimental value of 8.7$\pm$0.5 eV.~\cite{97CP-Bernas}
We see that the PBE functional predicts a band gap 1.8 eV smaller than HSE, while the original DeePKS yields negligible enhancement. However, both DeePKS-ES(s) and DeePKS-ES(u) models predict the band gap with variances less than 0.1 eV.
Based on the above, we conclude that both DeePKS-ES(s) and DeePKS-ES(u) effectively improve the accuracy of the original DeePKS model in predicting energy levels of water systems.

\subsection{SCF Convergence of DeePKS-ES}

While the DeePKS-ES(s) model is more accurate for specific systems, its ability to work with other systems that have different chemical environments is uncertain.
On the other hand, the DeePKS-ES(u) model is a bit less accurate than the DeePKS-ES(s) but should be readily applied to a wide variety of water systems, including both molecular and liquid forms.

\begin{figure}
  \includegraphics[width=8cm]{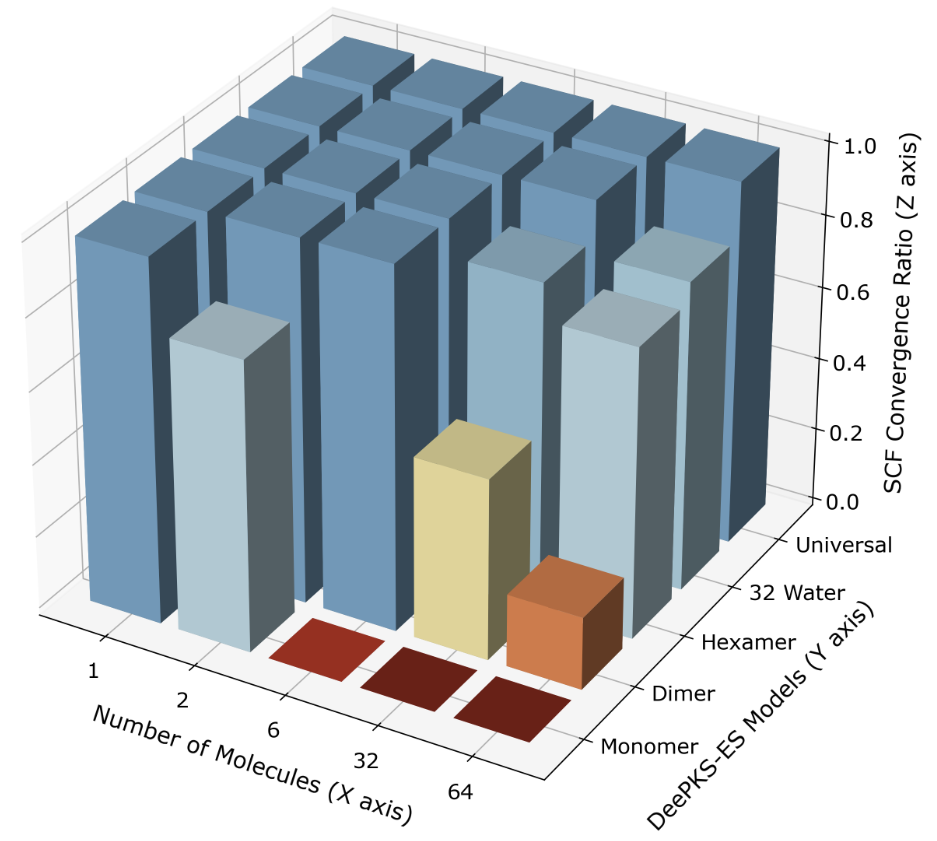}
  \caption{(Color online)
      SCF convergence ratios in solving Kohn-Sham equations for five different water systems with different DeePKS models. 
      The $X$-axis denotes five water systems including 1 (water monomer), 2 (water dimer), 6 (water hexamer) water molecules, as well as liquid water with 32 and 64 water molecules.
      Here the liquid water system comprising 64 molecules is absent from all of the training datasets.
      The $Y$-axis categorizes five different DeePKS-ES models. The labels ``Monomer," ``Dimer," ``Hexamer," and ``32 Water" refer to models uniquely trained on their respective water cluster systems, while ``Universal" denotes a universal model, which utilizes the training data from the previous four systems. 
      The $Z$-axis quantifies the SCF convergence ratio.
  }\label{conv_ratio}
\end{figure}

The SCF convergence is an important property of the DeePKS models, and we find the two models exhibit different behaviors.
In detail, Fig.~\ref{conv_ratio} shows the SCF convergence ratio of five different models when applied to five different water systems. The five models include one DeePKS-ES(u) model and four DeePKS-ES(s) models trained for water monomer, water dimer, water hexamer, liquid water with 32 molecules. The five testing systems are water monomer, water dimer, water hexamer, liquid water with 32 molecules, and liquid water with 64 molecules.
Here the SCF convergence ratio is defined as the percentage of successful convergent configurations out of 20 test configurations for each system.

Interestingly, our studies reveal that the individually trained DeePKS models possess robust SCF convergence capabilities for smaller molecular systems compared to their training systems. This suggests potential convergence issues when applying these DeePKS models to larger systems. 
For instance, the DeePKS-ES(s) model, initially trained on a water monomer, fails to converge for systems comprising more than two water molecules. 
Similarly, the model trained from water dimer results in a rapid decrease of SCF convergence rate when applied to large systems.
Notably, the model trained from liquid water with 32 molecules maintains convergence across all tested molecular configurations, but the convergence rate drops to 85\% for the liquid water system containing 64 molecules.
We note that during SCF calculations, the DeePKS-ES(s) models typically require 18 to 20 iterations for convergence, more than the 10 iteration steps by the PBE functional.

Contrastingly, the DeePKS-ES(u) model consistently outperforms others in SCF convergence across varied systems, including the liquid water systems with 64 water molecules, which are not included in the training data. 
Regarding the convergence steps, the number is around 15 for both 32- and 64-molecule liquid water systems, slightly fewer than those needed by the DeePKS-ES(s) models.
This performance indicates that the universal DeePKS-ES(u) model has acquired the capability to accommodate diverse chemical interactions in water systems.


\subsection{Overall Performance of DeePKS-ES}

\begin{figure*}
  \includegraphics[width=16cm]{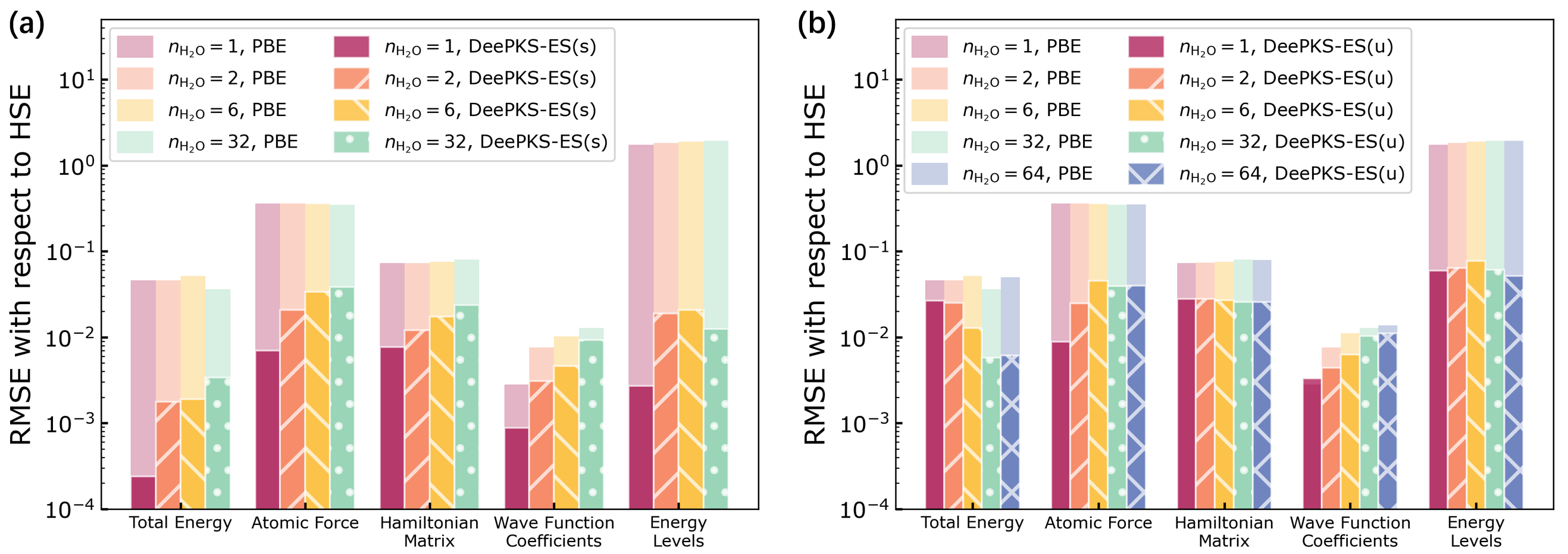}
  \caption{(Color online)
  RMSEs of total energy, atomic forces, Hamiltonian matrix, wave function coefficients, and energy levels from the PBE and DeePKS-ES models with respect to HSE across various systems are compared. 
  RMSE is defined in Eq.~\eqref{eq:RMSE-begin}-\eqref{eq:RMSE-end}.
  (a) The RMSE for both the PBE and separately optimized DeePKS-ES models, denoted as DeePKS-ES(s), when benchmarked against the HSE for systems with water monomer (1 molecule), water dimer (2 molecules), and water hexamer (6 water molecules) and a periodic arrangement with 32 water molecules, is detailed.
  (b) A similar RMSE contrast between the PBE and the generalized DeePKS-ES model, labeled DeePKS-ES(u), in relation to HSE for water molecular configurations of 1, 2, 6 and periodic systems with 32 and 64 molecules, is presented. The 64-molecule periodic system was not included in the model's training dataset.
  }\label{DeePKS-ES_RMSE}
\end{figure*}

To further check the overall accuracy of the DeePKS-ES(s) and DeePKS-ES(u) models, Figs.~\ref{DeePKS-ES_RMSE}(a) and (b) respectively illustrate the RMSEs of total energy, atomic forces, Hamiltonian matrix, wave function coefficients, and energy levels for a few water systems, which include the water monomer, water dimer, water hexamer and liquid water (32 water molecules) systems.
Detailed values are shown in Supplementary Table S5.
In addition, Fig.~\ref{DeePKS-ES_RMSE}(b) adds an additional liquid water system (64 water molecules), which is absent from the training data.
The RMSEs for each system are averaged over 20 test configurations.

Figure~\ref{DeePKS-ES_RMSE}(a) reveals that the DeePKS-ES(s) model consistently outperforms the PBE functional across the four evaluated systems. 
First, the RMSEs for the total energy in DeePKS-ES(s) models are notably lower than those of PBE. 
For instance, the RMSE for the water monomer is reduced to $2.4\times10^{-4}$ eV/atom and for the water dimer to $1.8\times10^{-3}$ eV/atom, compared to the PBE's RMSE of $4.6\times10^{-2}$ eV/atom for both systems.
Second, the DeePKS-ES(s) models achieve excellent improvements in RMSE for atomic forces, with reductions of more than an order of magnitude relative to PBE.
Third, for the Hamiltonian matrix, DeepPKS-L(s) models show substantial improvements over PBE, although this enhancement is not as significant as those observed in the previous two terms. In addition, the RMSE term of the wave function coefficients exhibits some improvement. 
Last, the energy level term exhibits the largest improvements, with the RMSE decreasing by almost two orders of magnitude compared to PBE.
Overall, we conclude that the RMSE increases with the number of molecules in the system, except for the energy levels in liquid water.
The increasing complexity posed by greater numbers of molecules indicates a higher challenge associated with training larger systems.

We subsequently validate the predictive accuracy of the DeePKS-ES(u) model.
The results shown in Fig.~\ref{DeePKS-ES_RMSE}(b) indicate that the universal model consistently surpassed PBE in all systems tested, implying its strong capability to deliver accurate results across various systems.
In addition, we also train a universal model based on the original DeePKS model.
The results regarding RMSEs are also shown in Supplementary Table S5, as well as those of the DeePKS-ES(s) models and the DeePKS-ES(u) model for comparison.
As expected, this model shows excellent accuracy for the total energy and atomic forces but little improvement over the electronic structural properties compared to the base functional PBE.

Within the five tested properties, the RMSE of total energy from the DeePKS-ES(u) decreases with the increase in the number of molecules in the system.
This trend may stem from larger systems featuring more training data, consequently enhancing deep neural network precision for the total energy.
Conversely, the RMSEs of DeePKS-ES(u) for the atomic forces and wave function coefficients gradually increase with a larger system, suggesting that the inherent complexity of learning these two terms in large systems.
Examining the PBE-predicted Hamiltonian matrix, the RMSE remains almost as a constant across various water systems.
Notably, unlike the previously discussed DeePKS-ES(s) models, the universal model learns the pattern, implying that the DeePKS-ES(u) model is fairly stable in prediciting the Hamiltonian matrix for water systems.
Regarding the wavefunction coefficients, the RMSEs between PBE and HSE are not as large as other terms, and we find the universal model offers slight enhancements over PBE.
The performance of the DeePKS-ES(u) model excels in terms of energy levels, exhibiting a notable decrease in RMSE compared to PBE.
In summary, DeePKS-ES(u) demonstrates superior performance, reducing the RMSE in atomic forces and energy levels by more than an order of magnitude when compared to PBE.

In particular, when compared to the liquid water system with 32 molecules, all of the five RMSEs from the liquid water system with 64 water molecules are similar, demonstrating the excellent transferability of the DeePKS-ES(u) model because the liquid system with 64 water molecules is absent from the training set.
Overall, we conclude that the universal model DeePKS-ES(u) shows excellent transferability for all the tested water systems.

\subsection{Efficiency of DeePKS-ES}

\begin{figure}
  \includegraphics[width=8cm]{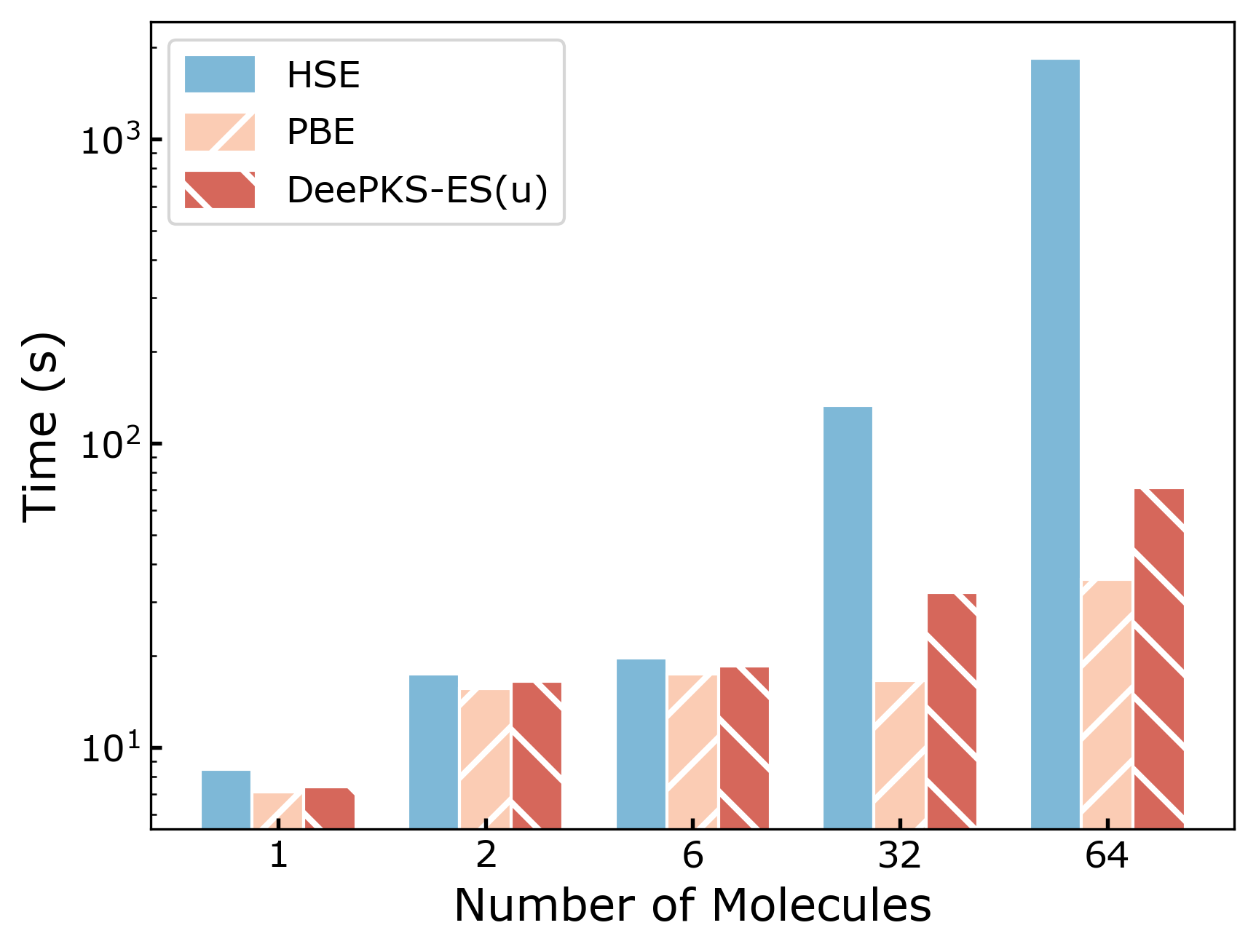}
  \caption{(Color online)
 Comparison of execution time for a single electronic iteration in DFT calculations across molecular systems comprising a water monomer (1 molecule), a water dimer (2 molecules), a water hexamer (6 molecules), and liquid water configurations of 32 and 64 molecules. We evaluate the efficiency of three methods, incorporating the HSE and PBE exchange-correlation functionals, as well as the DeePKS-ES(u) approach. Our computational setup employs 12 CPU cores powered by an Intel(R) Xeon(R) Platinum 8260 CPU @ 2.40GHz.
  }\label{scf_time}
\end{figure}

The computational efficiency of DeePKS models is shown in Fig~\ref{scf_time}, which displays the average time for an electronic step during the SCF calculations utilizing the PBE, HSE, and DeePKS-ES methods.
The average time is computed by dividing the total computational time by the number of iterative steps.
These computations were executed on a system with 12 CPU cores of the Intel(R) Xeon(R) Platinum 8260 CPU operating at 2.40 GHz.
Due to its universality, the DeePKS-ES(u) model is employed to highlight the computational demands of DeePKS models across various systems; other DeePKS models display similar efficiency.
We find the DeePKS-ES(u) model significantly outperforms the HSE functional in efficiency, only modestly exceeding the computational time of using the PBE functional.
For instance, completing one electronic step in a liquid configuration of 64 water molecules demanded approximately 4195 seconds with HSE and only 25 seconds with PBE. 
Meanwhile, the DeePKS-ES(u) model completed the same job in about 58 seconds, roughly twice as long as PBE. This marks a huge advancement by nearly two orders of magnitude over HSE while delivering comparable precision.
Moreover, as the system grows in complexity with increasing water molecule counts, the computational costs of the DeePKS method grow much slower than those of HSE and are similar to those of PBE.
This demonstrates the DeePKS-ES model has the potential to study larger sizes of systems.
To summarize, the DeePKS-ES(u) model shares similar accuracy with HSE but is more efficient.

\section{Conclusions}

In this work, we have developed an enhanced Deep Kohn-Sham (DeePKS) method, named DeePKS-ES, to address the limitations of the original DeePKS method in predicting electronic structure properties. 
The original DeePKS method uplifts the accuracy of a base exchange-correlation functional by employing a computationally efficient neural-network-based correction term, which is trained against a higher-level exchange-correlation functional. 
It is effective in predicting total energy and atomic forces but is unable to accurately predict electronic structure properties such as the Hamiltonian matrix and energy levels.
To overcome this, we introduced the DeePKS-ES method, which incorporates the Hamiltonian matrix and its eigenvalues and eigenvectors into the loss function during training. 
This innovation enabled the method to not only maintain high accuracy in predicting total energy and atomic forces but also to accurately predict electronic structure properties.

We applied the DeePKS-ES method to various water systems, including water monomers, dimers, hexamers, and liquid water.
We constructed both individual and universal models, employing PBE as the base functional and HSE as the target functional.
The individual models DeePKS-ES(s) were trained on specific systems, while the universal model DeePKS-ES(u) was trained on a diverse dataset encompassing all water systems, including molecular and liquid ones.
They demonstrated high accuracy with respect to the target XC functional HSE.
For instance, in the water monomer system, the DeePKS-ES(u) model showed a total energy difference of only $1.06\times10^{-5}$ eV per atom while achieving remarkable accuracy in electronic structure properties. 
The maximum difference in the Hamiltonian matrix elements was just 0.056 Ry, and the band gap error was measured at $8.39\times10^{-2}$ eV.
Regarding the DOS of liquid water systems, both the DeePKS-ES(s) and DeePKS-ES(u) models produced results that closely matched that from the HSE functional.
This represented a significant enhancement over the original DeePKS model and the base functional PBE.

Moreover, the DeePKS-ES(u) model exhibited excellent transferability and robustness across different water systems, including those not included in the training data. 
This universal model maintained high accuracy in predicting electronic structure properties and showed significant improvements over the PBE functional, especially in energy levels, where the RMSE was reduced by over an order of magnitude.

In addition to the superior accuracy, the DeePKS-ES models offered significant computational efficiency. 
The universal DeePKS-ES model achieved comparable precision to the HSE functional while being nearly two orders of magnitude faster, requiring only slightly more computational time than PBE.
This efficiency makes the DeePKS-ES model highly suitable for studying larger and more complex systems with high accuracy.

In summary, the DeePKS-ES method represents a significant advancement in predicting electronic structure properties with high accuracy and efficiency. 
It effectively bridges the gap between low-accuracy, efficient functionals and high-accuracy, computationally expensive methods, making it a powerful tool for studying large-scale systems.
Currently, the DeePKS-ES method is limited to calculations with a single k point, leaving the implementation of multi-K point calculations for the future.
Further optimizing the model for broader applications and exploring its potential in other material systems are also future directions.

$\\$
{\bf Acknowledgements}
$\\$
The authors thank Han Wang, Qi Ou, Yixiao Chen, and Linfeng Zhang for helpful discussions. We thank the electronic structure team (from AI for Science Institute, Beijing) for improving the ABACUS package from various aspects. The work of X.L, R.L. and M.C. is supported by the National Natural Science Foundation of China under Grant No. 11988102 and 12135002. All of the numerical simulations were performed on the High-Performance Computing Platform of CAPT.

\bibliography{reference}
\end{document}


\title{Supplemental Material for ``A Deep Learning Framework for the Electronic Structure of Water: Towards a Universal Model"}

\author{Xinyuan Liang}
\affiliation{Academy for Advanced Interdisciplinary Studies, Peking University, Beijing, 90871, P. R. China}
\affiliation{HEDPS, CAPT, College of Engineering, Peking University, Beijing, 100871, P. R. China}
\author{Renxi Liu}
\affiliation{Academy for Advanced Interdisciplinary Studies, Peking University, Beijing, 90871, P. R. China}
\affiliation{HEDPS, CAPT, College of Engineering, Peking University, Beijing, 100871, P. R. China}
\author{Mohan Chen}
\thanks{$^\ast$Corresponding author. Email:mohanchen@pku.edu.cn}
\affiliation{AI for Science Institute, Beijing 100080, P. R. China}
\affiliation{Academy for Advanced Interdisciplinary Studies, Peking University, Beijing, 90871, P. R. China}
\affiliation{HEDPS, CAPT, College of Engineering, Peking University, Beijing, 100871, P. R. China}

\maketitle

\section{Finite Difference Test for Atomic Forces}

\begin{figure}[H]
\centering
  \includegraphics[width=12cm]{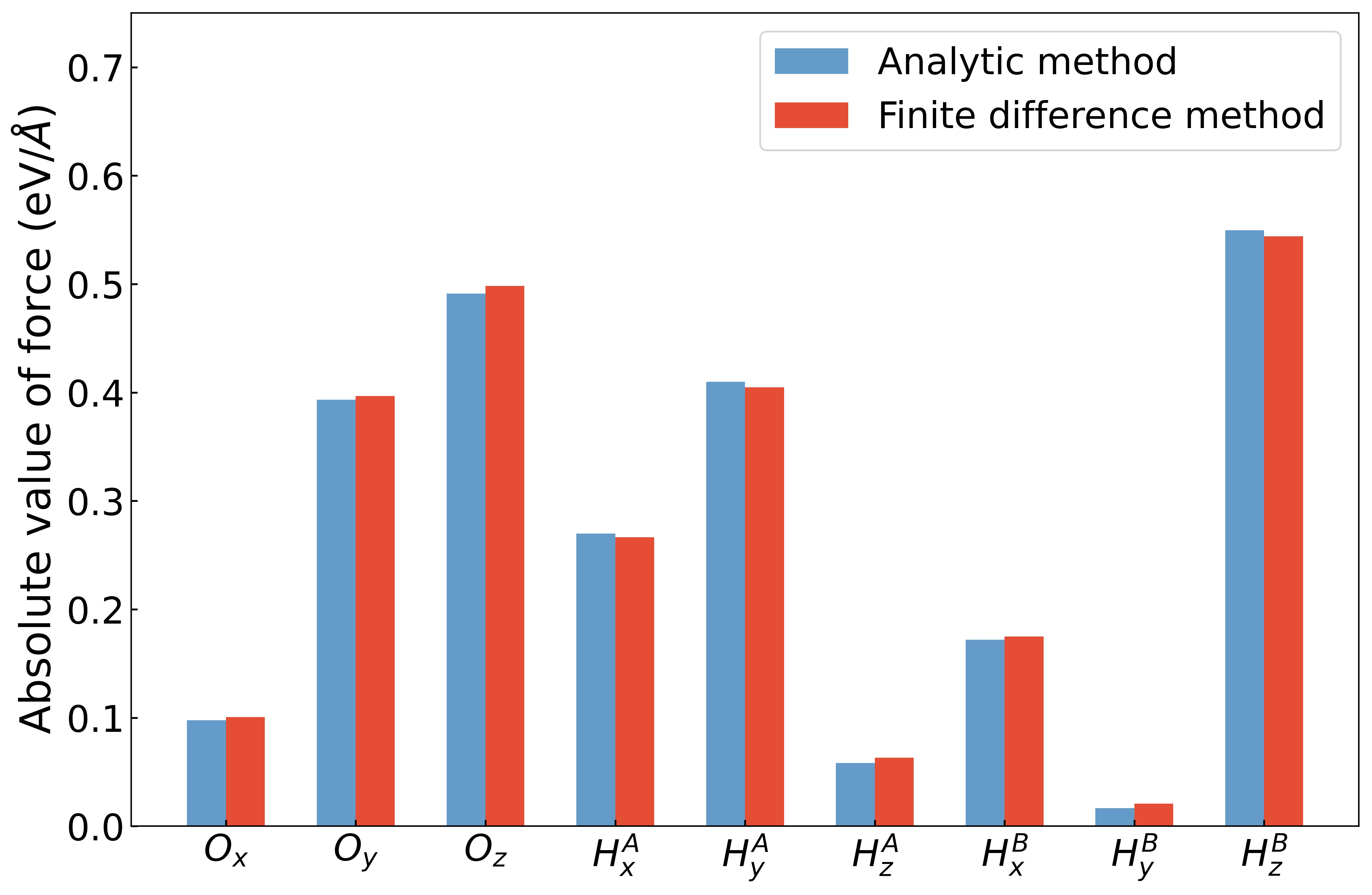}
  \caption{
  Finite difference test for the atomic forces of the DeePKS model in the water monomer system.
  The figure shows the absolute values of the forces along the three directions of one O and two H atoms obtained by the two calculation methods.
  The blue one is obtained by the analytic method, while the red one is obtained by the finite difference method.
  }
  \label{finite_difference}
 \end{figure}

Fig.~\ref{finite_difference} shows the absolute values of the forces on water monomer in three directions calculated by two different calculation methods.
The analytic method uses the calculation formula mentioned in the main text to calculate the atomic forces.
In addition, the finite difference method to compute the atomic forces takes the formula of
\begin{equation}
\label{eq:finite_difference}
\begin{aligned}
F^i_{\alpha}\left(\mathbf{R}^1, \ldots, \mathbf{R}^N\right)=&\frac{\partial}{\partial R_\alpha^i} E\left(\mathbf{R}^1, \ldots, \mathbf{R}^N\right)\\
\approx& \frac{E\left(\ldots, \mathbf{R}^i+\Delta R \hat{\mathbf{e}}_\alpha, \ldots\right)-E\left(\ldots, \mathbf{R}^i-\Delta R \hat{\mathbf{e}}_\alpha, \ldots\right)}{2 \Delta R}.
\end{aligned}
\end{equation}
$R^i_{\alpha}$ and $F^i_{\alpha}$ represent the coordinate and force on the $i$-th atom in the $\alpha$ direction, respectively.
$\mathbf{e}_\alpha$ represents the unit vector in the $\alpha$ direction, which is chosen as 0.001 Bohr in Fig.~\ref{finite_difference}.
The figure demonstrates a notable similarity between the atomic forces calculated via the analytical method and those obtained through the finite difference method. 
The similarity indicates the accuracy of energy and atomic force formulas, highlighting a consistent relationship between them.

\begin{figure}[H]
\centering
  \includegraphics[width=12cm]{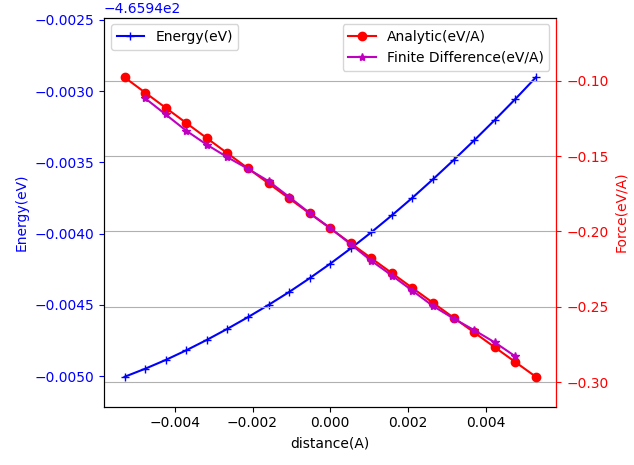}
  \caption{
    Finite difference test of the DeePKS model for the oxygen atom in the water monomer along the x-direction.
    The horizontal axis illustrates how far the oxygen atom deviates from its initial arrangement along the x direction.
    The blue line indicates the total energy of the corresponding configuration.
    The red and purple lines represent the atomic forces obtained by the analytic and finite difference methods.
  }
  \label{finite_difference_O_x}
 \end{figure}

Fig.~\ref{finite_difference_O_x} shows a more detailed finite difference test for the oxygen atom in the water monomer along the x direction.
The horizontal axis represents the deviation of the oxygen atom from its original configuration in the x direction, with a distance of 0.001 Bohr between each two adjacent points.
The blue and red lines indicate the total energy and atomic force by the analytic method for the corresponding configuration, respectively.
The purple line represents the atomic forces obtained by the finite difference method.
As in Eq.~\ref{eq:finite_difference}, it is obtained by dividing the energy difference of the two adjacent configurations by $2 \Delta R$, with $\Delta R$ being 0.001 Bohr.
In all these configurations, the forces obtained by the analytical method and the finite difference method show satisfactory similarity.

\newpage
\section{Performance of DeePKS-ES Models}
\subsection{RMSE of Total Energy and Atomic Forces in Water Monomer}
\begin{table}[H]
    \caption{RMSE of total energy, atomic forces given by PBE, DeePKS, DeePKS-ES(s), and DeePSK-ES(u) with respect to the HSE functional in the water monomer system.
    $n_{H_2O}$ and $n_{occ}$ represent the number of water molecules and the occupied wave function, respectively.
    }
    \label{table:sm_monomer_ef}
        \begin{center}
    \begin{tabular}{cc|cccc|cccc}
    \hline
    \hline
    \multicolumn{2}{c|}{RMSE} & \multicolumn{4}{c|}{Total Energy(eV/atom)} & \multicolumn{4}{c}{Atomic Forces(eV/\AA)} \\
    \hline
        $n_{H_2O}$ & $n_{occ}$ & PBE & DeePKS & DeePKS-ES(s) & DeePKS-ES(u) & PBE & DeePKS & DeePKS-ES(s) & DeePKS-ES(u)   \\
    \hline
    1 & 4 & 0.04604 & 0.00001 & 0.00026 & 0.02673 & 0.35765 & 0.00604 & 0.00614 & 0.01177 \\
    \hline
    \hline
    \end{tabular}
    \end{center}
\end{table}

\newpage

\subsection{Difference of Energy Levels in Water Monomer}
\begin{table}[htb]
\label{table:sm_band_monomer}
\caption{
Energy levels given by HSE as well as the difference in energy levels given by PBE, DeePKS, DeePKS-ES(s), and DeePKS-ES(u) with respect to HSE in water monomer system.
``Index" represents the index of energy levels.
This is the tabular data of Fig. 4(c) in the main text.}
    \begin{center}
    \begin{tabular}{c|cccccccc}
        \hline
        \hline
        Index & 1 & 2 & 3 & 4 & 5 & 6 & 7 & 8\\
        \hline
        HSE & -2.807e+01 & -1.463e+01 & -1.062e+01 & -8.554e+00 & 1.426e+00 & 4.873e+00 & 1.076e+01 & 1.145e+01\\
        PBE-HSE & 2.829e+00 & 1.465e+00 & 1.524e+00 & 1.489e+00 & -6.562e-01 & -9.327e-01 & -9.323e-01 & -7.493e-01\\
        DeePKS-HSE & 3.142e+00 & 1.587e+00 & 1.741e+00 & 1.702e+00 & -5.515e-01 & -7.249e-01 & -8.150e-01 & -6.369e-01\\
        DeePKS-ES(s)-HSE & 2.800e-03 & -2.000e-04 & 2.800e-03 & -2.070e-03 & -4.000e-04 & 2.560e-03 & -1.100e-03 & 2.100e-03\\
        DeePKS-ES(u)-HSE & 7.700e-03 & -6.030e-02 & -8.570e-02 & -7.829e-02 & 5.590e-03 & 9.831e-02 & 2.000e-03 & 1.870e-02\\
        \hline
        \hline
    \end{tabular}
    \end{center}
\end{table}


\newpage
\subsection{Difference of Energy Levels in Water Dimer}
\begin{table}[htb]
\renewcommand{\arraystretch}{1.5}
\caption{Energy levels given by HSE as well as the difference in energy levels given by PBE, DeePKS, DeePKS-ES(s) and DeePKS-ES(u) with respect to HSE in water dimer system.
``Index" represents the index of energy levels.
This is the tabular data of Fig. 5(a) in the main text.}
    \begin{center}
    \begin{tabular}{c|cccccccc}
        \hline
        \hline
        Index & 1 & 2 & 3 & 4 & 5 & 6 & 7 & 8\\
        \hline
        HSE & -2.830e+01 & -2.788e+01 & -1.495e+01 & -1.429e+01 & -1.076e+01 & -1.068e+01 & -8.687e+00 & -8.578e+00\\
        PBE-HSE & 2.839e+00 & 2.811e+00 & 1.487e+00 & 1.427e+00 & 1.502e+00 & 1.535e+00 & 1.488e+00 & 1.488e+00\\
        DeePKS-HSE & 2.928e+00 & 2.899e+00 & 1.271e+00 & 1.234e+00 & 1.396e+00 & 1.405e+00 & 1.189e+00 & 1.189e+00\\
        DeePKS-ES(s)-HSE & -2.860e-02 & -9.700e-03 & 4.700e-03 & 1.330e-02 & 2.200e-03 & 7.400e-03 & -4.900e-03 & 4.890e-03\\
        DeePKS-ES(u)-HSE & 9.300e-03 & 4.800e-03 & -5.500e-02 & -6.370e-02 & -8.310e-02 & -8.180e-02 & -7.589e-02 & -8.656e-02\\
        \hline
        \hline
        Index & 9 & 10 & 11 & 12 & 13 & 14 & 15 & 16\\
        \hline
        HSE & 9.799e-01 & 1.601e+00 & 4.384e+00 & 5.016e+00 & 1.055e+01 & 1.066e+01 & 1.123e+01 & 1.132e+01\\
        PBE-HSE & -6.445e-01 & -6.726e-01 & -9.264e-01 & -9.317e-01 & -9.115e-01 & -9.280e-01 & -9.643e-01 & -7.560e-01\\
        DeePKS-HSE & -6.578e-01 & -6.819e-01 & -9.518e-01 & -9.449e-01 & -8.545e-01 & -8.728e-01 & -9.002e-01 & -8.027e-01\\
        DeePKS-ES(s)-HSE & -2.983e-02 & -1.107e-02 & 1.273e-02 & 5.990e-03 & -1.340e-02 & -4.570e-02 & -9.800e-03 & 2.050e-02\\
        DeePKS-ES(u)-HSE & -1.426e-02 & 6.820e-03 & 5.013e-02 & 1.101e-01 & -1.260e-02 & 1.330e-02 & 3.010e-02 & 2.390e-02\\
        \hline
        \hline
    \end{tabular}
    \end{center}
\end{table}

\newpage
\subsection{Difference of Energy Levels in Water Hexamer}
\begin{table}[H]
\renewcommand{\arraystretch}{1.5}
\caption{Energy levels given by HSE as well as the difference in energy levels given by PBE, DeePKS, DeePKS-ES(s) and DeePKS-ES(u) with respect to HSE in water hexamer system.
``Index" represents the index of energy levels.
This is the tabular data of Fig. 5(b) in the main text.}
    \begin{center}
    \begin{tabular}{c|cccccccc}
        \hline
        \hline
        Index & 1 & 2 & 3 & 4 & 5 & 6 & 7 & 8\\
        \hline
        HSE & -2.844e+01 & -2.813e+01 & -2.795e+01 & -2.775e+01 & -2.726e+01 & -2.708e+01 & -1.470e+01 & -1.458e+01\\
        PBE-HSE & 2.849e+00 & 2.843e+00 & 2.827e+00 & 2.863e+00 & 2.827e+00 & 2.821e+00 & 1.511e+00 & 1.478e+00\\
        DeePKS-HSE & 3.110e+00 & 3.124e+00 & 3.116e+00 & 3.136e+00 & 3.102e+00 & 3.106e+00 & 1.288e+00 & 1.266e+00\\
        DeePKS-ES(s)-HSE & 3.390e-02 & 1.750e-02 & 2.240e-02 & -1.900e-03 & 2.900e-03 & -5.400e-03 & -2.500e-03 & 2.130e-02\\
        DeePKS-ES(u)-HSE & -4.910e-02 & -5.950e-02 & -5.690e-02 & -7.250e-02 & -4.810e-02 & -5.410e-02 & -6.650e-02 & -5.650e-02\\
        \hline
        \hline
        Index & 9 & 10 & 11 & 12 & 13 & 14 & 15 & 16\\
        \hline
        HSE & -1.443e+01 & -1.432e+01 & -1.402e+01 & -1.357e+01 & -1.169e+01 & -1.147e+01 & -1.135e+01 & -1.068e+01\\
        PBE-HSE & 1.461e+00 & 1.499e+00 & 1.518e+00 & 1.485e+00 & 1.529e+00 & 1.523e+00 & 1.525e+00 & 1.492e+00\\
        DeePKS-HSE & 1.248e+00 & 1.277e+00 & 1.285e+00 & 1.269e+00 & 1.398e+00 & 1.395e+00 & 1.393e+00 & 1.346e+00\\
        DeePKS-ES(s)-HSE & 2.220e-02 & 9.700e-03 & -5.300e-03 & 1.430e-02 & 4.800e-03 & 4.400e-03 & 2.900e-03 & 3.300e-03\\
        DeePKS-ES(u)-HSE & -6.030e-02 & -7.090e-02 & -8.550e-02 & -6.000e-02 & -8.260e-02 & -7.910e-02 & -8.450e-02 & -7.610e-02\\
        \hline
        \hline
        Index & 17 & 18 & 19 & 20 & 21 & 22 & 23 & 24\\
        \hline
        HSE & -1.016e+01 & -9.743e+00 & -8.945e+00 & -8.865e+00 & -8.779e+00 & -8.378e+00 & -8.350e+00 & -8.022e+00\\
        PBE-HSE & 1.467e+00 & 1.462e+00 & 1.408e+00 & 1.437e+00 & 1.418e+00 & 1.413e+00 & 1.478e+00 & 1.453e+00\\
        DeePKS-HSE & 1.313e+00 & 1.283e+00 & 1.123e+00 & 1.115e+00 & 1.114e+00 & 1.111e+00 & 1.127e+00 & 1.115e+00\\
        DeePKS-ES(s)-HSE & 8.500e-03 & 1.240e-03 & 7.560e-03 & 9.480e-03 & 3.369e-02 & 4.090e-03 & -8.110e-03 & 4.770e-03\\
        DeePKS-ES(u)-HSE & -7.750e-02 & -8.067e-02 & -7.260e-02 & -8.758e-02 & -5.529e-02 & -8.833e-02 & -9.377e-02 & -8.473e-02\\
        \hline
        \hline
        Index & 25 & 26 & 27 & 28 & 29 & 30 & 31 & 32\\
        \hline
        HSE & 4.669e-01 & 1.617e+00 & 1.655e+00 & 2.895e+00 & 3.033e+00 & 3.677e+00 & 5.532e+00 & 6.087e+00\\
        PBE-HSE & -5.699e-01 & -6.520e-01 & -6.461e-01 & -7.773e-01 & -7.793e-01 & -8.039e-01 & -9.643e-01 & -9.722e-01\\
        DeePKS-HSE & -5.420e-01 & -6.307e-01 & -6.287e-01 & -7.818e-01 & -7.868e-01 & -8.203e-01 & -1.015e+00 & -1.016e+00\\
        DeePKS-ES(s)-HSE & -5.712e-02 & -4.011e-02 & -1.231e-02 & 3.401e-02 & 3.687e-02 & 5.604e-02 & -2.657e-02 & -5.120e-03\\
        DeePKS-ES(u)-HSE & -1.099e-01 & -8.384e-02 & -6.045e-02 & 1.120e-02 & 6.230e-03 & 4.274e-02 & -1.052e-01 & -8.357e-02\\
        \hline
        \hline
        Index & 33 & 34 & 35 & 36 & 37 & 38 & 39 & 40\\
        \hline
        HSE & 6.518e+00 & 6.699e+00 & 7.250e+00 & 7.584e+00 & 8.137e+00 & 8.392e+00 & 9.747e+00 & 9.981e+00\\
        PBE-HSE & -1.002e+00 & -9.624e-01 & -1.030e+00 & -9.744e-01 & -8.308e-01 & -7.848e-01 & -9.000e-01 & -9.095e-01\\
        DeePKS-HSE & -1.049e+00 & -1.009e+00 & -1.066e+00 & -1.034e+00 & -7.759e-01 & -8.104e-01 & -8.023e-01 & -8.199e-01\\
        DeePKS-ES(s)-HSE & -2.628e-02 & -2.430e-03 & -2.577e-02 & -1.280e-02 & -5.285e-02 & -3.658e-02 & -3.434e-02 & -5.070e-02\\
        DeePKS-ES(u)-HSE & -1.152e-01 & -1.177e-01 & -1.722e-01 & -1.285e-01 & -1.032e-01 & -8.824e-02 & -5.568e-02 & -8.549e-02\\
        \hline
        \hline
        Index & 41 & 42 & 43 & 44 & 45 & 46 & 47 & 48\\
        \hline
        HSE & 1.068e+01 & 1.112e+01 & 1.118e+01 & 1.170e+01 & 1.225e+01 & 1.282e+01 & 1.326e+01 & 1.507e+01\\
        PBE-HSE & -8.919e-01 & -8.456e-01 & -8.474e-01 & -9.103e-01 & -9.148e-01 & -1.045e+00 & -1.064e+00 & -1.126e+00\\
        DeePKS-HSE & -8.275e-01 & -8.410e-01 & -8.463e-01 & -8.689e-01 & -8.729e-01 & -9.299e-01 & -9.203e-01 & -9.552e-01\\
        DeePKS-ES(s)-HSE & -6.200e-03 & 1.900e-02 & 2.000e-04 & 9.500e-03 & 6.300e-03 & -1.360e-02 & -7.400e-03 & 7.800e-03\\
        DeePKS-ES(u)-HSE & -3.240e-02 & -3.280e-02 & -4.700e-02 & -7.830e-02 & -5.860e-02 & -6.390e-02 & -2.680e-02 & -3.800e-03\\
        \hline
        \hline
    \end{tabular}
    \end{center}
\end{table}

\newpage
\subsection{RMSEs Given by Universal Models}
\begin{table}[H]
\renewcommand{\arraystretch}{1.5}
\setlength{\tabcolsep}{1.8mm}
\label{rmse_all}
\caption{Root mean square errors (RMSE) of total energy, atomic force, Hamiltonian matrix, wave function coefficients, and energy levels given by PBE, DeePKS(u), DeePKS-ES(u) and DeePKS-ES(s) with respect to the HSE functional in different systems.
See the main text for the definition of the RMSEs.
$n_{H_2O}$ represents the number of water molecules in the system.
This contains the data of Figs. 8(a) and (b) in the main text.
}

        \begin{center}
    \begin{tabular}{c|c|ccccc}
    \hline
    \hline
    RMSE & $n_{H_2O}$ & 1 & 2 & 6 & 32 & 64 \\
        \hline
        \multirow{3}{*}{Total Energy(eV/atom)}        & PBE & 0.04560 & 0.04567 & 0.05152 & 0.03627 & 0.04976\\
                                                        & DeePKS(u) & 0.00138 & 0.00144 & 0.00754 & 0.00032 & 0.01136\\
                                                        & DeePKS-ES(u) & 0.02673 & 0.02523 & 0.01272 & 0.00584 & 0.00609\\
                                                        & DeePKS-ES(s) & 0.00024 & 0.00176 & 0.00191 & 0.00338 & - \\
        \hline
        \multirow{3}{*}{Atomic Force(eV/\AA)}        & PBE & 0.36140 & 0.35842 & 0.35677 & 0.34540 & 0.34944\\
                                                        & DeePKS(u) & 0.04386 & 0.04321 & 0.03519 & 0.03063 & 0.02863\\
                                                        & DeePKS-ES(u) & 0.00888 & 0.02481 & 0.04506 & 0.03910 & 0.03997\\
                                                        & DeePKS-ES(s) & 0.00697 & 0.02066 & 0.03387 & 0.03827 & - \\
        \hline
        \multirow{3}{*}{Hamiltonian matrix(Ry)} & PBE & 0.07263 & 0.07319 & 0.07584 & 0.07971 & 0.07924\\
                                                        & DeePKS(u) & 0.08212 & 0.08275 & 0.08562 & 0.08957 & 0.08903\\
                                                        & DeePKS-ES(u) & 0.02781 & 0.02779 & 0.02702 & 0.02578 & 0.02573\\
                                                        & DeePKS-ES(s) & 0.00767 & 0.01199 & 0.01730 & 0.02372 & - \\
        \hline
        \multirow{3}{*}{Wave Function Coefficients}   & PBE & 0.00280 & 0.00761 & 0.01110 & 0.01281 & 0.01371\\
                                                        & DeePKS(u) & 0.00442 & 0.01480 & 0.02002 & 0.02007 & 0.01886\\
                                                        & DeePKS-ES(u) & 0.00329 & 0.00442 & 0.00629 & 0.01028 & 0.01115\\
                                                        & DeePKS-ES(s) & 0.00088 & 0.00308 & 0.00462 & 0.00932 & - \\
        \hline
        \multirow{3}{*}{Energy levels(eV)}        & PBE & 1.73111 & 1.81081 & 1.87768 & 1.93902 & 1.92940\\
                                                        & DeePKS(u) & 1.90060 & 1.98716 & 2.04740 & 2.06661 & 2.05691\\
                                                        & DeePKS-ES(u) & 0.05962 & 0.06407 & 0.07746 & 0.06102 & 0.05152\\
                                                        & DeePKS-ES(s) & 0.00272 & 0.01882 & 0.02076 & 0.01246 & - \\
        \hline
        \hline
    \end{tabular}
    \end{center}

\end{table}

\bibliography{SM_ref}